\documentclass[english,aps,prb,amsmath,nofootinbib,longbibliography]{revtex4-2}
\usepackage[T1]{fontenc}
\usepackage[latin9]{inputenc}
\usepackage{babel}
\usepackage{float}
\usepackage{bm}
\usepackage{dsfont}
\usepackage{amsmath}
\usepackage{amssymb}
\usepackage{graphicx}
\usepackage{esint}
\usepackage[unicode=true,pdfusetitle,
 bookmarks=true,bookmarksnumbered=false,bookmarksopen=false,
 breaklinks=false,pdfborder={0 0 1},backref=false,colorlinks=false]
 {hyperref}
\hypersetup{
 pdfborderstyle=}

\makeatletter

\newcommand{\lyxmathsym}[1]{\ifmmode\begingroup\def\b@ld{bold}
  \text{\ifx\math@version\b@ld\bfseries\fi#1}\endgroup\else#1\fi}

\usepackage{babel}
\usepackage{dsfont}
\usepackage{braket}
\DeclareMathOperator{\Tr}{Tr}
\usepackage{bbold}
\usepackage[compat=1.1.0]{tikz-feynman}

\makeatother

\begin{document}
\title{Magnetization nutation in magnetic semiconductors: \\
 Effective spin model with anisotropic RKKY exchange interaction}
\author{H. Kachkachi}
\affiliation{Université de Perpignan Via Domitia, Laboratoire PROMES-CNRS (UPR
8521), Rambla de la Thermodynamique, Tecnosud, 66100 Perpignan, FRANCE.}
\date{\today}
\begin{abstract}
We demonstrate that the magnetization in magnetic semiconductors exhibits
nutational motion when subjected to an external magnetic field. This
behavior originates from the splitting of the conduction-electron
band which induces anisotropic, distance-dependent exchange coupling
between localized spins. 

To investigate this phenomenon, we examine a general system that includes
both charge and spin degrees of freedom, characteristic of a magnetic
semiconductor. This system is composed of two subsystems: (1) a gas
of noninteracting conduction electrons and (2) a ferromagnetic array
of localized spins, coupled through the Vonsovskii (\textit{sd}) local
interaction. The entire system is subject to external electrical and
magnetic disturbances. Through the Feynman-Schwinger formalism, we
integrate out the faster (Grassmann) charge degrees of freedom associated
with the conduction electrons to obtain the effective Hamiltonian
for the localized spins in the form of an XXZ spin model. We then
provide general analytical formulas for the corresponding anisotropic
exchange couplings, expressed in both Fourier and direct spaces, as
functions of the effective field that induces conduction-band splitting.
As a practical application of this general formalism, we investigate
the dynamics of the net magnetic moment within the effective spin
model. We demonstrate, analytically and numerically, that the anisotropy
of the distance-dependent exchange coupling induces nutational magnetization
oscillations at significantly higher frequencies than the typical
precession, which is also modified accordingly. Analytical expressions
for the modified precession and low-lying nonuniform modes have been
derived and compared with the numerical solution of the system of
coupled Landau-Lifshitz equations for the localized spins. Our results
indicate that the frequency of the nutation modes decreases while
their amplitude increases as the band-splitting field intensifies.
Moreover, we show that the leading correction to the localized spin
gyromagnetic factor $g_{S}$, due to the polarization of conduction
electrons in an external magnetic field, is linear in the \textit{sd}
coupling. 

We hope this study will motivate further research into nutational
phenomena in magnetic semiconductors, possibly resulting in improvements
in the accurate control of magnetization dynamics within spin-based
electronic devices.
\end{abstract}
\maketitle

\section{Introduction}

Semiconductors and magnetic materials play an essential role in modern
electronics. In electronic and optical semiconductors, the charge
degrees of freedom are used to process information, whereas in magnetic
materials the spin is utilized to store information. One of the main
ideas that have been driving the developments of the last few decades
is to combine the properties of both materials for possible spin-electronic
applications with enhanced functionalities\citep{Prinz_sci90,WolfEtAl_sci01}.
This combination has been achieved in diluted magnetic semiconductors
(DMS) by doping a nonmagnetic semiconductor with magnetic elements
\citep{FurdynaKossut_ap}. The advent of spintronics based on the
manipulation of spin polarized currents has triggered an active search
for ferromagnetic semiconductors (FMS) and the discovery of long range
ferromagnetism beyond room temperature. Indeed, the discovery of ferromagnetism
in doped semiconductors with high transition temperatures has led
to a renewed interest in DMS \citep{OhnoEtAl_prl92,MatsukuraEtAl_Buschow02,SinghEtAl_prb03,JungwirthETAl_RevModPhys.78.809,DietlEtAl_ieee07}.
Besides their potential applications, magnetic semiconductors (MS)
have attracted considerable attention owing to the fundamental mechanism
and nature of their ferromagnetic state and the underlying magnetic
cooperative phenomena arising from the new spin degrees of freedom.
Likewise, magnetic materials such as EuO and EuS have been actively
investigated for decades because they exhibit a high degree of symmetry
and isotropy as well as short-range magnetic interaction, making them
simple model spin systems \citep{DietrichEtAl_prb75,MaugerGoddart_PR86,MaugerMills_PhysRevB.28.6553,whiwoo68pr,whiwoo70prb},
and for their widely recognized potential applications \citep{Dietl_sst02,KatayamaEtAl_re12,OrlovEtAl_re12,Hasan_arxiv24}.

A branch of spintronics applications focuses on information storage
on magnetic media by reversing the magnetization direction using various
means, such as an external magnetic field initially applied opposite
to the magnetization state. The efficiency of this (damped) switching
process is limited by the macroscopic relaxation time of the magnetization
which is on the order of 100 ps. More efficient switching processes,
achieving ultra-fast magnetization reversal, can be obtained using
picosecond magnetic field pulses perpendicular to the magnetization
direction \cite{GerritsEtAl_prb06,barros11prb}. In general, the dynamics
of magnetization in a magnetically ordered material is governed by
precession. However, if the latter is somehow disturbed so that the
rotation axis is no longer aligned with the angular momentum, an additional
motion appears known as \emph{magnetic nutation}, similar to what
occurs in the case of a mechanical gyroscope when an external force
tilts its rotation axis away from the direction of the gravity field.
For decades, magnetization nutation was ignored as being a spurious
effect on top of precession. However, wide attention has recently
been drawn towards this phenomenon owing partly to the recent experimental
observations of nutational oscillation at THz frequencies in thin
films~\citep{LiEtAl_prb15, *NeerajEtAl_nat21, *Unikandanunni_prl22, *DeEtAl_arxiv24}.
In fact, spin nutation was investigated a few decades ago in NMR~\citep{Torrey49pr},
optical resonance~\citep{hoctan68prl} and EPR~\citep{verfes73jcp,atkinsEtal74cpl,fedoruk02jas}.
It was then predicted in Josephson junctions~\citep{zhufra06jpcm,franzhu08njp,franson08nanotech,PhysRevB.71.214520,PhysRevLett.92.107001}
and was later developed using various approaches based on relativistic
quantum mechanics, first principles~\citep{mondaletal17prb,monopp_jpcm20},
and electronic structure theory~\citep{PhysRevLett.108.057204,PhysRevB.84.172403,kiktat_prb15,thoningetal17sp,chengetal17prb}.
More recent theoretical investigations deal with the magnetization
nutation in metallic ferromagnets as being induced by magnetic inertia
due to the nonadiabatic contribution of the environment degrees of
freedom\citep{kiktat_prb15,MakhlufEtAl_apl20,ThiNico_epjb21}. Magnetic
inertia has also been introduced within a macroscopic approach~\citep{ciorneietal11prb,oliveetal12apl,doi:10.1063/1.4921908,oliweg16jpcm}
through an extended version of the Landau-Lifshitz-Gilbert (LLG) equation~\citep{lanlif35,gilbert56phd}.
Recently, it has been shown \cite{basvenkac_prb18,AdamsEtAl_prb24_110}
that spin misalignment induced by surface effects in nanoscale magnetic
systems trigger nutational oscillations of the magnetization at several
frequencies in the GHz-THz range. There are also studies of nutation-like
excitations in systems of coupled vortices \cite{LeeEtAl_jap17,GarGus_nm23}
and elongated nano-elements \cite{UrzOcam_j3m20}.

One of the most important features of MS (FMS and DMS) is that their
magnetic properties, such as the ferromagnetic phase transition and
spin-excitation dynamics can be controlled by the interaction between
the spins of itinerant carriers and the spins of the localized magnetic
ions, the so called \emph{sd} coupling. This provides us with a handle
to control their magnetic properties by light through its action on
the charge degrees of freedom. Indeed, several recent experiments
have used time-resolved optical techniques to demonstrate electrical
control of coherent electron-spin dynamics in nonmagnetic semiconductor
hetero-structures. Since this exploits variations in the spin-orbit
interaction due to compositional gradients, the energy of these effects
in conventional semiconductors, however, is on the $\mu$eV scale and is
manifested as variations in the spin precession frequency in the GHz
range. In contrast, THz electron spin precession frequencies can be
easily achieved in MS due to the \emph{sd} exchange interaction. This
coupling leads to large spin splittings in the conduction band states
on the order of 10-100 meV \citep{AwschalomSmarth02,MyersEtAl_prb05},
thus matching the THz regime of spin nutation. This is the reason
we believe that it is quite relevant and timely to study magnetic
nutation in MS and, in particular, DMS whose outstanding properties
and functionalities stem from the complex structure of their valence
band and the effect of long-range carrier-mediated exchange (RKKY)
interaction. More precisely, it is important to investigate their
low-lying energy states which contribute to both equilibrium ($M_{s}$
and $T_{\mathrm{C}}$) and nonequilibrium magnetic properties (FMR
and relaxation).

\medskip{}

\textbf{Objectives and plan of the paper}: 

One of the objectives of this work is to demonstrate that magnetic
nutation occurs in magnetic semiconductors, driven by the resulting
anisotropic, distance-dependent long-range exchange coupling between
the localized spins. To achieve this, we proceed as follows: In the
next section, we present a model of a magnetic semiconductor consisting
of a subsystem of conduction electrons coupled to the subsystem of
localized spins. We then derive the effective spin Hamiltonian for
the localized spins using the path integral formalism to integrate
out the Grassmann variables associated with the fast charge degrees
of freedom.

Next, we solve the system of Landau-Lifshitz equations for the localized
spins and analyze the time evolution of the net magnetic moment, highlighting
the precessional and nutational oscillations, which depend on the
band-splitting field. In the subsequent section, we derive analytical
expressions for the precession and nutation frequencies and compare
them with the numerical results. We also compute the amplitude of
the nutational mode as a function of the band-splitting field.

Appendix A details the derivation of the effective spin Hamiltonian,
while Appendix B provides the explicit calculation of the long-range
effective exchange couplings in both Fourier and direct spaces.

\section{\label{sec:Hamiltonian}Model Hamiltonian and hypotheses}

A crystal becomes a conductor when free electrons and holes appear
in it. The model proposed by Vonsovsky in 1971 {[}see the review by
Irkhin \cite{Irkhin_pmm2010} and references therein{]} for a magnetic
semiconductor (MS) presumes that all electrons in the crystal can
be subdivided into the localized ones that form partially filled ionic
$d$- or $f$-shells and the de-localized ones occupying $s$-type
states. Since the former are localized and the latter are de-localized
(carriers), this model is termed the $cl$ model. This is also sometimes
called the $sd$-model.

Accordingly, we consider a magnetic semiconductor as being composed
of two subsystems: 1) the subsystem of conduction electrons and 2)
the subsystem of localized spins. The two subsystems are coupled by
the \emph{sd} interaction of the conduction electron spin ${\bf s}_{i}$
with the localized spins $\bm{S}_{i}$. The corresponding contributions
to the total Hamiltonian are described below.

\paragraph{Conduction electrons}

The Hamiltonian of the conduction electrons is given by

\begin{eqnarray}
\mathcal{H}_{e} & = & \mathcal{H}_{\mathrm{K}}+\mathcal{H}_{{\rm Z}}^{\left(s\right)}\label{eq:eHam-System}
\end{eqnarray}
where the first term is the electron kinetic energy,\emph{ 
\begin{equation}
\mathcal{H}_{\mathrm{K}}=\eta\sum_{\left\langle i,j\right\rangle }\sum_{\sigma}C_{i,\sigma}^{+}C_{j,\sigma}\label{eq:eHam-Kinetic}
\end{equation}
}with $C_{i,\sigma}^{+}$ and $C_{i,\sigma}$ being the creation and
annihilation operators of the electron and $\eta$ the nearest-neighbor
hopping energy. We will consider the simplest case of a one-band model
for which the electrons are in the Bloch state with quasi-momentum
$\bm{k}$ and spin $\sigma$ {[}See Ref. \cite{takahashi_carrier_2010}{]}.

In Eq. (\ref{eq:eHam-System}), $\mathcal{H}_{{\rm Z}}^{\left(s\right)}$
is the Zeeman energy of the electrons in the external (DC) magnetic
field $\bm{H}_{0}$, \emph{i.e. }$\mathcal{H}_{{\rm Z}}^{\left(s\right)}=-g_{e}\mu_{\mathrm{B}}\bm{H}_{0}\cdot\sum_{i}\bm{s}_{i}$.
In the present approach, we will use the representation of the conduction
electron spin in terms of the creation and annihilation operators,
\emph{i.e.} ${\bf s}_{i}=\left(\hbar/2\right)\sum_{\alpha,\beta}C_{i,\alpha}^{+}\bm{{\bf \sigma}}_{\alpha\beta}C_{i,\beta}$,
with $\bm{{\bf \sigma}}$ being the vector of Pauli matrices. Therefore,
upon performing the Fourier transformation 
\begin{equation}
C_{\alpha}\left({\bf r}\right)=\frac{1}{\sqrt{V}}\sum_{{\bf k}}e^{-i{\bf k}.\bm{r}}\;C_{{\bf k}\alpha},C_{\alpha}^{\dagger}\left({\bf r}\right)=\frac{1}{\sqrt{V}}\sum_{{\bf k}}e^{i{\bf k}.\bm{r}}\;C_{{\bf k}\alpha}^{+},\label{eq:FT_CC+}
\end{equation}
the total Hamiltonian of the subsystem of conduction electrons becomes

\begin{equation}
\mathcal{H}_{e}=\sum_{\bm{k}}E_{\bm{k}}\sum_{\sigma}C_{\bm{k}\sigma}^{+}C_{\bm{k}\sigma}-\sum_{\alpha,\beta}\sum_{{\bf k}}C_{{\bf k}\alpha}^{+}\left(\bm{\xi}\cdot{\bf \bm{\sigma}}_{\alpha\beta}\right)C_{{\bf k}\beta}\label{eq:eHam_FT}
\end{equation}
where 
\begin{align*}
E_{\bm{k}} & =z\eta t_{\bm{k}},\quad t_{\bm{k}}=\frac{1}{z}\sum_{\bm{a}}e^{i\bm{k}\cdot\bm{a}}=\frac{1}{d}\sum_{\alpha=x,y,z}\cos\left(ak_{\alpha}\right),\\
\bm{\xi} & =\frac{1}{2}\gamma_{s}\mu_{\mathrm{B}}\bm{H}_{0}.
\end{align*}
$d$ is the dimension of space, $z$ the coordination number and $a$
the lattice spacing.

\paragraph{Localized spins}

The subsystem of localized spins subjected to an external magnetic
field $\bm{H}_{0}$ can be described by the (Heisenberg) Hamiltonian\cite{nagaev_spin_1974,nagaev83mir,MaugerGoddart_PR86,sutarto_euo_2009}
\begin{equation}
\mathcal{H}_{{\rm S}}=-J\sum_{\left\langle i,j\right\rangle }\bm{S}_{i}\cdot\bm{S}_{j}-g_{S}\mu_{\mathrm{B}}\bm{H}_{0}\cdot\sum_{i}\bm{S}_{i}\label{eq:Ham-LocalizedSpins}
\end{equation}
where $g_{S}$ is the gyromagnetic factor of the local spin $\bm{S}_{i}$.

We note that the superexchange term in Eq. (\ref{eq:Ham-LocalizedSpins})
is initially absent in the case of a DMS and the exchange coupling
between localized spins appears only upon doping with magnetic ions.
There is a huge literature on this topic {[}see the review articles\cite{SatoEtAl_RevModPhys.82.1633,Dietl_nm10,KALITA2023116201}
and references therein{]}. In ferromagnetic semiconductors, such as
EuO and EuS, this superexchange term couples the spins of Eu atoms.
In these compounds, the nearest-neighbor exchange coupling ($J_{1}$)
between localized spins is the strongest coupling. Indeed, the next-nearest-neighbor
coupling ($J_{2}$), inferred from specific heat and neutron inelastic
scattering diffusion experiments reported in Refs. \onlinecite{MaugerGoddart_PR86,sutarto_euo_2009},
is very small ($J_{2}/J_{1}\sim0.2$).

\paragraph{Electrical disturbance}

In addition to the magnetic disturbance represented by the Zeeman
energy in Eqs. (\ref{eq:eHam-System}) and (\ref{eq:Ham-LocalizedSpins}),
we may include an electric or a charge disturbance in the form\citep{kimetal73prb}
\begin{align}
\mathcal{H}_{\mathrm{Elec}} & =\sum_{\sigma}\intop d{\bf r}\,\Phi_{{\rm pl}}\left({\bf r},t\right)\,C_{\sigma}^{+}\left({\bf r}\right)C_{\sigma}\left({\bf r}\right),\label{eq:ElecPot-on-Electrons}
\end{align}
where $\Phi_{{\rm pl}}\left({\bf r},t\right)=-\left|e\right|\varphi_{s}\left({\bf r},t\right)$.

The latter contribution is added for completeness anticipating on
future applications. For example, it will be interesting to investigate
the coupling between a plasmonic nanostructure with an FMS\citep{verkac2024phys}.
Accordingly, we can consider a system composed of an assembly of metallic
(Au, Ag, etc) nano-elements (of a few tens of nanometers in diameter)
deposited on top of an FMS, such as EuO or EuS. A monochromatic light
beam shed on the nano-elements at appropriate frequencies excites
plasmonic modes in the latter leading to the creation of hot spots
with enhanced electric field. The latter penetrates (with some attenuation)
into the FMS and alters the behavior of the conduction electrons which
in turn, through their coupling to the localized spins, affect the
excitations spectrum of the latter. Here, the effect of the top layer
of nano-elements is regarded as an (external) excitation electromagnetic
field, which leads to new contributions in the effective Hamiltonian
of localized spins. So the potential in Eq. (\ref{eq:ElecPot-on-Electrons})
may stem from the electric field radiated by the assembly of metallic
nanoparticles or film excited at their plasmonic resonance by an incident
electromagnetic wave.

\paragraph{Total Hamiltonian of the magnetic semiconductor}

As discussed earlier, the exchange interaction between the localized
(ion) spin ${\bf S}_{i}$ at site $\bm{r}_{i}$ and the conduction-electron
spin $\bm{s}_{i}$ is given by

\begin{equation}
\mathcal{H}_{sd}=-\lambda\sum_{i}{\bf S}_{i}\cdot{\bf s}_{i}=-\frac{\lambda}{2}{\displaystyle \sum}_{i}\sum_{\alpha,\beta}C_{i,\alpha}^{+}\left({\bf S}_{i}\cdot\bm{\sigma}_{\alpha\beta}\right)C_{i,\beta}\label{eq:Ham-sdCoupling}
\end{equation}
where $\lambda$ is the $sd$ coupling. This is a local interaction,
see Refs. \cite{nagaev_spin_1974,nagaev83mir,Irkhin_pmm2010,GajKos_spr11}
for extensive discussions.

In our derivation of the effective spin Hamiltonian by integrating
out the charge degrees of freedom, presented in Appendix \ref{sec:Effective-action},
we drop the spin Hamiltonian $\mathcal{H}_{{\rm S}}$ in Eq. (\ref{eq:Ham-LocalizedSpins})
because it does not involve the electron gas. Therefore, the total
Hamiltonian for the charge degrees of freedom, comprising the (free)
conduction electron energy (\ref{eq:eHam_FT}), the \emph{sd }coupling
(\ref{eq:Ham-sdCoupling}) and the coupling of the electrons to the
external electrical potential (\ref{eq:ElecPot-on-Electrons}), reads
\begin{align}
\mathcal{H} & =\sum_{{\bf k},\alpha}E_{{\bf k}}C_{{\bf k}\alpha}^{\dagger}C_{{\bf k}\alpha}-\sum_{\alpha,\beta}\sum_{{\bf k}}C_{{\bf k}\alpha}^{+}\left(\bm{\xi}\cdot{\bf \bm{\sigma}}_{\alpha\beta}\right)C_{{\bf k}\beta}+\frac{1}{V}\sum_{{\bf k},{\bf p}}\sum_{\alpha,\beta}C_{{\bf k}\alpha}^{+}\Upsilon_{\alpha\beta}\left({\bf k}-{\bf p},t\right)C_{{\bf p},\beta},\label{eq:fms-in-plas}
\end{align}
where 
\begin{equation}
\Upsilon_{\alpha\beta}\left({\bf k},t\right)\equiv\Phi_{{\rm pl}}\left({\bf k},t\right)\delta_{\alpha\beta}-\frac{\lambda}{2}\left({\bf S}_{{\bf k}}\left(t\right)\cdot\bm{\sigma}_{\alpha\beta}\right).\label{eq:TotalPot-on-Electrons}
\end{equation}

Due to the Zeeman splitting under the effect of the external magnetic
field and the internal molecular field, induced by localized spins
${\bf S}$, the perturbed energies of the free carriers are 
\begin{align}
E_{\bm{k},\sigma}\equiv E_{\sigma}\left(\bm{k}\right) & =E_{\bm{k}}-\hbar\frac{\sigma}{2}\left(g\mu_{\mathrm{B}}H_{0}+\lambda\left\langle S^{z}\right\rangle \right).\label{eq:Energy-perturbed}
\end{align}

A few remarks are now in order. 
\begin{itemize}
\item According to Eq. (\ref{eq:Energy-perturbed}), the (rigid) band splitting
between the spin-up and spin-down bands is caused by the effective
field $\Xi$ comprising the external magnetic field and the internal
molecular field, \emph{i.e. 
\begin{equation}
\Xi=\xi+\frac{\lambda}{2}\left\langle S^{z}\right\rangle .\label{eq:BandSplitting}
\end{equation}
}This splitting parameter will appear in all subsequent calculations
via the ratio $\Xi/\epsilon_{\mathrm{F}}$, where $\epsilon_{\mathrm{F}}$
is the Fermi energy of the electron gas. In the mean-field picture,
the external magnetic field aligns the localized spins (magnetic ions),
which in turn act on the carriers (conduction electrons) via the ion-carrier
interaction $\lambda$. However, in practice, in (doped) magnetic
semiconductors, such as Eu$_{1-x}$Gd$_{x}$O and Ga$_{1-x}$Mn$_{x}$,
the main contribution to $\Xi$ is attributed to the second term \citep{jenmac_ox1991},
\emph{i.e.} to the \emph{sd} coupling $\lambda$. For example, in
an external magnetic field of 1T, in an MS with $S=5/2-7/2$ and $\epsilon_{\mathrm{F}}\simeq0.4\,\mathrm{eV}$,
$\xi/\epsilon_{\mathrm{F}}\simeq10^{-4}$ while $\Xi/\epsilon_{\mathrm{F}}\simeq10^{-1}-1$.
In the textbook \cite{GajKos_spr11}, there is a detailed discussion,
and in Ref. \cite{BeaDan_prb10} a table of values, of the coupling
$\lambda$ and the splitting it produces in the conduction and valence
bands in various MS. 
\item For the derivation of the effective spin Hamiltonian {[}Appendix \ref{sec:Effective-action}{]},
we will adopt the path-integral approach for which one chooses the
grand-canonical ensemble. Indeed, the density of electrons in a real
material fluctuates around the average $N/V$, which is held fixed
by the chemical potential $\mu$ as the temperature is varied. Then,
if we adopt a parabolic dispersion for the conduction electrons, we
write $E_{k}=\hbar^{2}k^{2}/2m-\mu$. 
\item In Ref. \cite{wesselinowa83PSS} it was shown that the spin-wave mode
is underdamped for low temperatures and overdamped near $T_{c}$,
see the review \cite{MaugerGoddart_PR86} for a detailed discussion.
Likewise, at low temperature the electronic damping $\gamma_{{\rm el}}^{\sigma}$
is extremely small, it is quadratic in the coupling $\lambda$ and
only increases as $T_{c}$ is approached. Consequently, in the present
work, we will ignore the damping $\gamma_{\bm{q}\sigma}$, since we
only consider the low temperature regime where spin-wave theory is
valid \citep{whiwoo70prb}. 
\end{itemize}

\section{\label{sec:ESH}Effective spin Hamiltonian: magnetization nutation}

In this section, we demonstrate that the magnetization of the underlying
(effective) spin subsystem of magnetic semiconductors, when subjected
to an external magnetic field, exhibits nutational motion. This motion
arises from the anisotropy between spin-up and spin-down conduction
bands, which induces anisotropic, distance-dependent exchange coupling. 

\subsection{Effective spin Hamiltonian: anisotropic RKKY exchange couplings}

To obtain the effective spin Hamiltonian for the magnetic semiconductor,
we assume, as hinted to earlier, that the motion of the conduction
electrons within the MS is much faster than the variations associated
with the fluctuations of the localized spins {[}see discussion in
the textbook \cite{nagaev83mir}{]}. Therefore, we can average over
the Grassmann variables associated with the conduction electrons using
the Feynman-Schwinger technique \citep{feynman_mathematical_1950,schwinger_gauge_1951}
and obtain the effective model for the localized spin variables. There
are several variants of such a procedure in the literature {[}see
\emph{e.g.}, Refs. \citep{negorl98,KonigEtAl_prl00,KonigEtAl_prb01,MudryWS2014}{]};
it is most efficiently done within the Lagrange formalism. In Appendix
\ref{sec:Effective-action}, we present the details of our more general
calculations including both electric and magnetic external excitations.
Considering the \emph{sd} coupling $\lambda$ as a small parameter
with respect to the band width, we perform a perturbation to $2^{\mathrm{nd}}$
order in $\lambda$. The whole procedure leads to the following effective
action for the subsystem of localized spins (focusing here only on
the external magnetic disturbance) 
\begin{quotation}
\begin{align}
\mathcal{A}_{{\rm eff}} & =-\frac{\lambda}{2V}\sum_{\bm{k}}\left[f_{\mathrm{FD}}\left(E_{\bm{k}}^{-}\right)-f_{\mathrm{FD}}\left(E_{\bm{k}}^{+}\right)\right]\left({\bf S}_{{\bf 0}}\cdot\bm{e}_{\xi}\right)\label{eq:Aeff-noPP}\\
 & +\frac{1}{2}\left(\frac{\lambda}{2V}\right)^{2}\sum_{\bm{p},\bm{k}}\frac{1}{\beta}\sum_{m}\left[{\bf S}\left(-k\right)\cdot{\bf S}\left(k\right)\right]\Bigg\{\frac{f_{\mathrm{FD}}\left(E_{\bm{p}}^{-}\right)-f_{\mathrm{FD}}\left(E_{\bm{p}+\bm{k}}^{+}\right)}{i\varpi_{m}+E_{\bm{p}}-E_{\bm{p}+\bm{k}}-2\Xi}+\left(\Xi\longleftrightarrow-\Xi\right)\Bigg\}\nonumber \\
 & +\frac{1}{2}\left(\frac{\lambda}{2V}\right)^{2}\sum_{\bm{k},\bm{p}}\frac{1}{\beta}\sum_{m}\left[{\bf S}\left(-k\right)\cdot\bm{e}_{\xi}\right]\left[{\bf S}\left(k\right)\cdot\bm{e}_{\xi}\right]\nonumber \\
 & \Bigg\{\frac{f_{\mathrm{FD}}\left(E_{\bm{p}}^{-}\right)-f_{\mathrm{FD}}\left(E_{\bm{p}+\bm{k}}^{-}\right)}{i\varpi_{m}+E_{\bm{p}}-E_{\bm{p}+\bm{k}}}-\frac{f_{\mathrm{FD}}\left(E_{\bm{p}}^{-}\right)-f_{\mathrm{FD}}\left(E_{\bm{p}+\bm{k}}^{+}\right)}{i\varpi_{m}+E_{\bm{p}}-E_{\bm{p}+\bm{k}}-2\Xi}+\left(\Xi\longleftrightarrow-\Xi\right)\Bigg\}\nonumber 
\end{align}
\end{quotation}
where $\bm{e}_{\xi}=\bm{\xi}/\xi$ is the verse of the applied magnetic
field, ${\bf S}_{{\bf 0}}=\sum_{i}\bm{S}_{i}$, $f_{\mathrm{FD}}\left(z\right)=1/\left(e^{\beta z}+1\right)$
is the Fermi-Dirac function, $E_{\bm{p}}^{\pm}=E_{\bm{p}}\pm\Xi$,
and $\varpi_{l}=2\pi l/\beta$ are the bosonic Matsubara frequencies
associated with the spin degrees of freedom, with $\beta=1/\left(k_{\mathrm{B}}T\right)$.
The presence of $\left(\Xi\longleftrightarrow-\Xi\right)$ in the
equations above and subsequent ones implies that the first term within
the curly brackets has to be duplicated upon changing the sign of
$\Xi$. The details are given in Appendix \ref{sec:Effective-action}.

The expression of the effective action above comprises the $1^{\mathrm{st}}$-order
contribution given by the first term and the $2^{\mathrm{nd}}$-order
contribution given by the remaining terms. In the absence of the magnetic
field ($\bm{\xi}=\bm{0}$), the $1^{\mathrm{st}}$-order contribution
vanishes and the $2^{\mathrm{nd}}$-order one reduces to

\begin{align}
\mathcal{A}_{{\rm eff}}^{\left(2\right)} & =\left(\frac{\lambda}{2V}\right)^{2}\sum_{\bm{k}}\left[\sum_{\bm{p}}\frac{1}{\beta}\sum_{m}\frac{f_{\mathrm{FD}}\left(E_{\bm{p}}\right)-f_{\mathrm{FD}}\left(E_{\bm{p}+\bm{k}}\right)}{i\varpi_{m}+E_{\bm{p}}-E_{\bm{p}+\bm{k}}}\right]\left({\bf S}_{-k}\cdot{\bf S}_{k}\right).\label{eq:2ndOrder-Aeff_RKKY}
\end{align}

This is the well known result from the RKKY theory, where the term
between square brackets brings an exchange coupling between localized
spins induced by their \emph{sd} coupling to conduction electrons.
The formula in Eq. (\ref{eq:Aeff-noPP}) is a non trivial extension
of the RKKY expression in the presence of an effective magnetic field.
This is the first original contribution of the present work.

In the more general situation with $\bm{\xi}\neq\bm{0}$, the result
in Eq. (\ref{eq:Aeff-noPP}) shows that the applied magnetic field
has several effects. To $1^{\mathrm{st}}$-order in the \emph{sd }coupling
$\lambda$, the magnetic field redefines the gyromagnetic factor $g_{S}$
of the localized spins. Indeed, the total Zeeman effect on the localized
spins $\bm{S}_{i}$ can be written as 
\[
\mathcal{H}_{\mathrm{Z}}=-g_{S}^{\mathrm{eff}}\left(\mu_{\mathrm{B}}\bm{H}_{0}\right)\cdot\sum_{i}\bm{S}_{i}
\]
with the effective gyromagnetic factor 
\begin{equation}
g_{S}^{\mathrm{eff}}\equiv g_{S}-\frac{\lambda}{2H_{0}}\frac{1}{V}\sum_{\bm{k}}\Big[f_{\mathrm{FD}}\left(E_{\bm{k}}^{-}\right)-f_{\mathrm{FD}}\left(E_{\bm{k}}^{+}\right)\Big]=g_{S}-\frac{\lambda}{4H_{0}}\frac{1}{V}\sum_{\bm{k}}\frac{\sinh\left(\beta\xi\right)}{\cosh\left(\beta E_{\bm{k}}^{+}/2\right)\cosh\left(\beta E_{\bm{k}}^{-}/2\right)}.\label{eq:GMF-eff}
\end{equation}

Expanding in\footnote{$\beta\xi=g_{e}\mu_{\mathrm{B}}H_{0}/k_{\mathrm{B}}T$. Then, using
$\mu_{\mathrm{B}}/k_{\mathrm{B}}\simeq0.672\,\mathrm{S.I.},$we see
that $\beta\xi\simeq0.0134$ for $T=50\,\mathrm{K}$ and a magnetic
field of $1\mathrm{T}$.} $\beta\xi\ll1$, this simplifies into 
\begin{align*}
g_{S}^{\mathrm{eff}} & \simeq g_{S}-\frac{\lambda\beta}{8}g_{e}\mu_{\mathrm{B}}\left[I_{1}\left(\beta\right)+\frac{\beta^{2}}{12}I_{2}\left(\beta\right)\xi^{2}\right]
\end{align*}
where 
\[
I_{1}\left(\beta\right)=\int\frac{d\bm{k}}{\left(2\pi\right)^{3}}\frac{1}{\cosh^{2}\left(\beta E_{\bm{k}}/2\right)},\quad I_{2}\left(\beta\right)=\int\frac{d\bm{k}}{\left(2\pi\right)^{3}}\frac{-1+3\tanh^{2}\left(\beta E_{\bm{k}}/2\right)}{\cosh^{2}\left(\beta E_{\bm{k}}/2\right)}.
\]
We have replaced the discrete sums over momenta to integrals using
$\frac{1}{V}\sum_{\bm{k}}F\left(\bm{k}\right)\rightarrow\int\frac{d\bm{k}}{\left(2\pi\right)^{3}}F\left(\bm{k}\right)$.

This shows that the correction to the gyromagnetic factor $g_{S}$
of the localized spin, due to the polarization of conduction electrons
in the external magnetic field, is linear in the $sd$ coupling $\lambda$.
This effect could be used to study the sign of the \emph{sd} coupling
in various materials \cite{MyersEtAl_prb05,balcerzak_pss06,moraesetal_sr19}.
This will be done in Ref. \cite{verkac2024phys}.

To $2^{\mathrm{nd}}$-order in the $sd$ coupling $\lambda$, the
applied magnetic field has a strong effect on the exchange coupling.
Indeed, from the result in (\ref{eq:Aeff-noPP}) we infer the contribution
of the \emph{sd} coupling to the effective spin Hamiltonian for the
localized spins ${\bf S}_{i}$ which we write as follows

\begin{align}
\mathcal{H}_{{\rm eff}} & =-\sum_{i,j}\mathcal{J}_{1}\left(\bm{r}_{ij}\right){\bf S}_{i}\cdot{\bf S}_{j}-\sum_{i,j}\mathcal{J}_{2}\left(\bm{r}_{ij}\right)\left({\bf S}_{i}\cdot\bm{e}_{\xi}\right)\left({\bf S}_{j}\cdot\bm{e}_{\xi}\right)\label{eq:Ham_Eff}
\end{align}
where 
\[
\mathcal{J}_{n}\left(\bm{r}_{ij}\right)\equiv\frac{1}{V}\sum_{\bm{k}}e^{-i\bm{k}\cdot\bm{r}_{ij}}\frac{1}{\beta}\sum_{m}\mathcal{J}_{n}\left(\bm{k},\omega\right),n=1,2
\]
with 
\begin{align}
\mathcal{J}_{1}\left(\bm{k},\omega,\Xi\right) & =-\frac{\lambda^{2}}{8}\frac{1}{V}\sum_{\bm{p}}\Bigg[\frac{f\left(E_{\bm{p}}^{-}\right)-f\left(E_{\bm{p}+\bm{k}}^{+}\right)}{\omega+E_{\bm{p}}^{-}-E_{\bm{p}+\bm{k}}^{+}}+\left(\Xi\longrightarrow-\Xi\right)\Bigg],\label{eq:Jey1}\\
\nonumber \\
\mathcal{J}_{2}\left(\bm{k},\omega,\Xi\right) & =-\frac{\lambda^{2}}{8}\frac{1}{V}\sum_{\bm{p}}\Bigg\{\frac{f\left(E_{\bm{p}}^{-}\right)-f\left(E_{\bm{p}+\bm{k}}^{-}\right)}{\omega+E_{\bm{p}}^{-}-E_{\bm{p}+\bm{k}}^{-}}+\left(\Xi\longrightarrow-\Xi\right)\Bigg\}-\mathcal{J}_{1}\left(\bm{k},\omega\right).\label{eq:Jey2}
\end{align}

Note that, in the case of an FMS, the contribution (\ref{eq:Ham_Eff})
would add to the spin Hamiltonian in Eq. (\ref{eq:Ham-LocalizedSpins}),
whereas in DMS the exchange coupling between localized spins is indirect
and is given solely by (\ref{eq:Ham_Eff}).

In the next section, we will study the magnetization dynamics of the
effective spin Hamiltonian (\ref{eq:Ham_Eff}) under a magnetic field
applied in the $z$ direction, \emph{i.e.} $\bm{e}_{\xi}=\bm{e}_{z}$.
Then, the Hamiltonian (\ref{eq:Ham_Eff}) can be rewritten in the
following XXZ form 
\begin{align}
\mathcal{H}_{{\rm eff}} & =-\sum_{i,j}\mathcal{J}_{\perp}\left(\bm{r}_{ij}\right)\left(S_{i}^{x}S_{j}^{x}+S_{i}^{y}S_{j}^{y}\right)-\sum_{i,j}\mathcal{J}_{\parallel}\left(\bm{r}_{ij}\right)S_{i}^{z}S_{j}^{z},\label{eq:XXZHam}
\end{align}
where 
\begin{equation}
\mathcal{J}_{\parallel}\left(\bm{k},\omega;\xi\right)=\mathcal{J}_{1}+\mathcal{J}_{2}=-\frac{\lambda^{2}}{8}\frac{1}{V}\sum_{\bm{p}}\Bigg\{\frac{f\left(E_{\bm{p}}^{-}\right)-f\left(E_{\bm{p}+\bm{k}}^{-}\right)}{\omega+E_{\bm{p}}^{-}-E_{\bm{p}+\bm{k}}^{-}}+\left(\Xi\longrightarrow-\Xi\right)\Bigg\}\label{eq:Jparal}
\end{equation}
and
\begin{equation}
\mathcal{J}_{\perp}\left(\bm{k},\omega;\xi\right)=\mathcal{J}_{1}\left(\bm{k},\omega\right)=-\frac{\lambda^{2}}{8}\frac{1}{V}\sum_{\bm{p}}\Bigg[\frac{f\left(E_{\bm{p}}^{-}\right)-f\left(E_{\bm{p}+\bm{k}}^{+}\right)}{\omega+E_{\bm{p}}^{-}-E_{\bm{p}+\bm{k}}^{+}}+\left(\Xi\longrightarrow-\Xi\right)\Bigg].\label{eq:Jperp}
\end{equation}

In our investigation of magnetic nutation in MS, we focus on the spatial
properties of the RKKY exchange couplings. These are defined by the
inverse Fourier transforms of the static limit ($\omega=0$) of the
couplings (\ref{eq:Jparal}) and (\ref{eq:Jperp}). Accordingly, in
Appendix \ref{sec:Effective-exchange-couplings}, we derive the following
explicit expressions for the exchange couplings:
\begin{align*}
\mathcal{J}_{\parallel}\left(\bm{k},0;\Xi\right) & =\frac{\lambda^{2}}{8}\left[\frac{\rho_{\mathrm{F}}^{\uparrow}}{2}\mathcal{F}\left(\frac{k}{2k_{\mathrm{F}}^{\uparrow}}\right)+\frac{\rho_{\mathrm{F}}^{\downarrow}}{2}\mathcal{F}\left(\frac{k}{2k_{\mathrm{F}}^{\downarrow}}\right)\right],\tag{\ref{eq:Jparal_mumu}}\\
\mathcal{J}_{\perp}\left(\bm{k},0;\Xi\right) & =\frac{\lambda^{2}}{8}\left[\frac{\rho_{\mathrm{F}}^{\uparrow}}{2}\Lambda^{\uparrow}\left(\Xi\right)\mathcal{F}\left(\Lambda^{\uparrow}\left(\Xi,\right)\frac{k}{2p_{\mathrm{F}}^{\uparrow}}\right)+\frac{\rho_{\mathrm{F}}^{\downarrow}}{2}\Lambda^{\downarrow}\left(\Xi\right)\mathcal{F}\left(\Lambda^{\downarrow}\left(\Xi,\right)\frac{k}{2p_{\mathrm{F}}^{\downarrow}}\right)\right]\tag{\ref{eq:Jperp-mu_v3}}
\end{align*}
where we have introduced the prefactor 
\begin{equation}
\Lambda^{\mu}\left(\Xi\right)=1+\frac{\Xi}{2\epsilon_{\mathrm{F}}^{\mu}}\left(\frac{k}{2k_{\mathrm{F}}^{\mu}}\right)^{-2}\label{eq:Jperp_prefactor}
\end{equation}
and 

\[
\mathcal{F}\left(x\right)=\frac{1}{2}+\frac{1-x^{2}}{4x}\log\left[\frac{x+1}{x-1}\right]
\]
is the usual Lindhard function.

We have introduced the usual (modified) Fermi momentum and energy
\begin{align*}
k_{\mathrm{F}}^{\mu} & =\sqrt{k_{\mathrm{F}}^{2}+\mu\frac{2m\Xi}{\hbar^{2}}}=k_{\mathrm{F}}\sqrt{1+\mu\tilde{\Xi}},\quad\epsilon_{\mathrm{F}}=\frac{\hbar^{2}k_{\mathrm{F}}^{2}}{2m},\\
\epsilon_{\mathrm{F}}^{\mu} & =\frac{\hbar^{2}\left(k_{\mathrm{F}}^{\mu}\right)^{2}}{2m}=\epsilon_{\mathrm{F}}+\mu\Xi,\mu=\pm1\,\mbox{(or \ensuremath{\uparrow\downarrow})},
\end{align*}
together with the density of states at the Fermi energy 
\[
\rho_{\mathrm{F}}^{\mu}=\frac{mk_{\mathrm{F}}^{\mu}}{\hbar^{2}\pi^{2}}=\rho_{\mathrm{F}}\sqrt{1+\mu\tilde{\Xi}},\qquad\tilde{\Xi}\equiv\frac{\Xi}{\epsilon_{\mathrm{F}}}.
\]
The density of states of a three dimensional parabolic spectrum (adopted
here) is $\rho\left(\epsilon\right)=\frac{\left(2m\right)^{3/2}}{2\pi^{2}\hbar^{3}}\sqrt{\epsilon}$
and $\rho_{\mathrm{F}}=\rho\left(\epsilon_{\mathrm{F}}\right)=\frac{mk_{\mathrm{F}}}{\hbar^{2}\pi^{2}}$.

The result (\ref{eq:Jperp-mu_v3}) coincides with that of Ref. \citep{jenmac_ox1991},
and the result in \citep{werwil_msp06} is recovered if $\Lambda^{\mu}\left(\Xi\right)\rightarrow 1$ (\emph{i.e.} for $\Xi\ll\epsilon_{\mathrm{F}}$).
For $\Xi=0$, we have
\[
\mathcal{J}_{\perp}\left(\bm{k},0;0\right)=\frac{\lambda^{2}}{8}\rho_{\mathrm{F}}\mathcal{F}\left(\frac{k}{2k_{\mathrm{F}}}\right).
\]

As discussed in \emph{e.g.} Ref. \citep{jenmac_ox1991}, to leading
order, $\mathcal{J}_{\perp}\left(\bm{k},0,\Xi\right)$ is not affected
by the external magnetic field and the induced extra polarization
of the conduction electrons may be accounted for in the Fermi wave
vector. In this case, $\mathcal{J}_{\perp}\left(\bm{k},0,\Xi\right)$
becomes 
\begin{align}
\mathcal{J}_{\perp}\left(\bm{k},0;\Xi\right) & \simeq\frac{\lambda^{2}}{8}\frac{\rho_{\mathrm{F}}}{2}\left[\mathcal{F}\left(\frac{k}{2k_{\mathrm{F}}^{\uparrow}}\right)+\mathcal{F}\left(\frac{k}{2k_{\mathrm{F}}^{\downarrow}}\right)\right].\label{eq:Jperp_ZeroField}
\end{align}

Comparing the expressions for $\mathcal{J}_{\parallel}$ and $\mathcal{J}_{\perp}$,
given in Eqs. (\ref{eq:Jparal_mumu}) and (\ref{eq:Jperp-mu_v3}),
respectively, we see that because of the factor $\Lambda^{\mu}\left(\Xi\right)$,
these effective exchange couplings are different, thus showing that
the splitting field $\Xi$ between the spin-up and spin-down bands
induces an anisotropy in the effective exchange coupling. Indeed,
if we drop the field dependence in the prefactors $\Lambda^{\mu}\left(\Xi\right)$
in Eq. (\ref{eq:Jperp-mu_v3}), we see that $\mathcal{J}_{\perp}\left(\bm{r}_{ij}\right)=\mathcal{J}_{\parallel}\left(\bm{r}_{ij}\right)$.
This implies that, at this level of approximation, there is no anisotropy
in the effective exchange coupling, as is the case in the standard
RKKY theory \citep{rudkit_pr54} $\mathcal{J}=-J_{0}\,\Phi\left(k/2k_{\mathrm{F}}\right)$,
with $J_{0}=\frac{\lambda^{2}k_{\mathrm{F}}^{3}\rho_{\mathrm{F}}}{2\pi}=\lambda^{2}mk_{\mathrm{F}}^{4}/\left(2\pi^{3}\hbar^{2}\right)$.
The function $\Phi$ is defined in Section \ref{subsec:Exchange-coupling-inDS}.

It was argued in Ref. \cite{YuLee_prb95} that, in DMS, the symmetry of the host crystal and thereby that of the wave functions may induce
a stronger directional dependence of the effective exchange coupling between the magnetic ions. Zhou et al. \cite{zhouetal_natphys10} studied the magnetic coupling between individual adatoms on platinum and also showed that the RKKY interaction is strongly direction-dependent.
In Ref. \cite{soorez_physe21} it was shown that in doped bilayer graphene, anisotropic RKKY interaction can be induced by a magnetic field. Similarly, the authors of Ref. \cite{zieneretal_prb04} present a derivation of the RKKY interaction in semiconductors accounting for the modulation of band edges and external fields. They also concluded that, in the presence of a Zeeman splitting of the bands, the RKKY Hamiltonian is no longer isotropic and can be written in the form of XXZ model.

In conclusion, the anisotropy of the RKKY exchange coupling, which arises in the present context from band spin splitting, can be disregarded when considering the system's ground state, \emph{i.e.}, when the localized spins (magnetic ions) and carrier spins are aligned parallel. However, when the carrier magnetization is tilted by an anisotropy field or an external magnetic field, as is the case here, the precession angles of the magnetizations differ. This disparity results in a nutational motion of the magnetization in the effective spin system, a phenomenon that will be demonstrated both numerically and analytically in the following sections.

\subsection{\label{subsec:Magnetic-nutation_LLE}Magnetic nutation: solution
of the Landau-Lifshitz equation }

In this section, we study the dynamics of the effective spin system of the localized spins $\bm{S}_{i},i=1,\ldots,\mathcal{N}$, whose
energy is given by the effective Hamiltonian $\mathcal{H}_{{\rm eff}}$
defined in Eq. (\ref{eq:XXZHam}), with the exchange couplings $\mathcal{J}_{\parallel}\left(\bm{r}_{ij}\right)$
and $\mathcal{J}_{\perp}\left(\bm{r}_{ij}\right)$ given in Eqs. (\ref{eq:Jparal-DirectSpace_v2},
\ref{eq:ExchangeCoupling_ww}). We define the net magnetic moment
$\bm{m}$ as $\bm{m}=\left(1/\mathcal{N}\right)\sum_{i=1}^{\mathcal{N}}\bm{S}_{i}$.

The dynamics of such a system is governed by the following system of coupled Landau-Lifshitz equations\footnote{We introduce the damped Landau-Lifshitz equation for generality, though we set the damping to zero in this work. A discussion of the effect of damping is given in the text.}
\begin{equation}
\frac{d\boldsymbol{S}_{i}}{d\tau}=\boldsymbol{S}_{i}\times\bm{h}_{{\rm eff},i}-\alpha\boldsymbol{S}_{i}\times\left(\boldsymbol{S}_{i}\times\bm{h}_{{\rm eff},i}\right),\label{eq:LLE}
\end{equation}
where $\alpha$ is the damping parameter and $\bm{h}_{{\rm eff},i}=-\delta\mathcal{H}_{{\rm eff}}/\delta\bm{S}_{i}$
the (normalized) local effective field acting on $\bm{S}_{i}$; $\tau$
is the reduced time $\tau=t/\tau_{s}$, where $\tau_{{\rm s}}=\mu_{\mathrm{a}}/\left(\gamma J\right)$
is the characteristic time of the system's dynamics; $\mu_{\mathrm{a}}$
is the magnetic moment per atom. Then, the frequency $\omega=2\pi f=2\pi/T$
is measured in units of $\tau_{{\rm s}}^{-1}$. The details of the
numerical procedure for studying the magnetic nutation in such many-spin
systems are described in Refs. \cite{basvenkac_prb18,AdamsEtAl_prb24_110}
and will not be repeated here. The main procedure uses the iterative
routine based on the $4^{\mathrm{th}}$-order Runge-Kutta scheme in
combination with the projection step $\boldsymbol{S}_{i}^{\nu+1}=\boldsymbol{S}_{i}^{\mathrm{RK4}}/\|\boldsymbol{S}_{i}^{\mathrm{RK4}}\|$
to preserve the constraint $\|\boldsymbol{S}_{i}\|=1$. We start with
an initial state where all (atomic) spins $\bm{S}_{i}$ are at some
angle with respect to the $z$ axis, materialized here by the direction
of the magnetic field, and then let the system freely evolve in time.
This means that we set to zero the damping parameter $\alpha$ so as to investigate the undamped excitation modes of the magnetic moment $\bm{m}$.

\begin{figure}[H]
\begin{centering}
\includegraphics[scale=0.6]{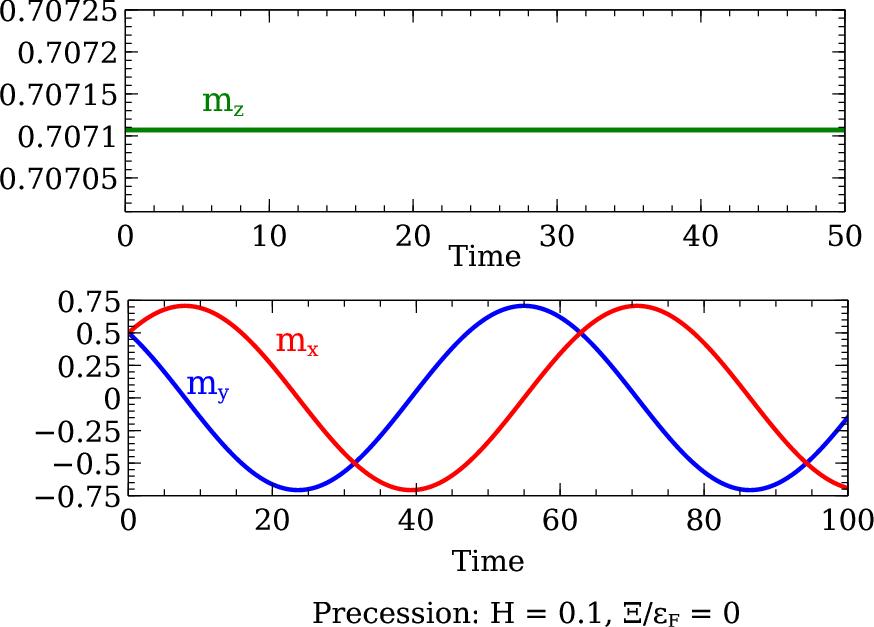}\includegraphics[scale=0.6]{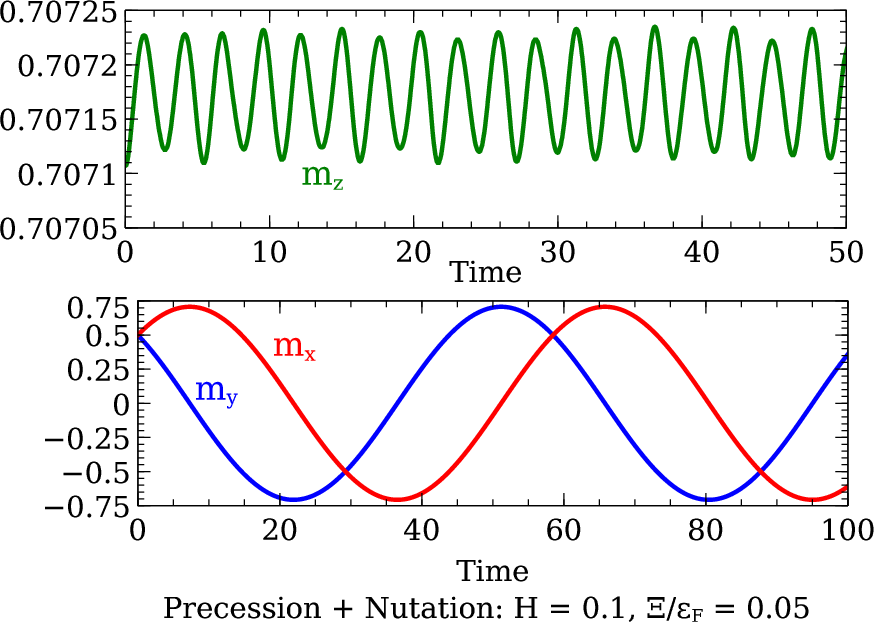}
\par\end{centering}
\caption{\label{fig:NumNut}Time evolution of the components of the (normalized) net magnetic moment of the subsystem of localized spins with $\Xi=0$ (left) and $\Xi\protect\neq0$ (right).}
\end{figure}

We solve the system of equations (\ref{eq:LLE}) to obtain the time
trajectory of each spin $\bm{S}_{i}$. Then, we compute the net moment
$\bm{m}$ and normalize it. The results are shown in Fig. \ref{fig:NumNut}
where we plot the time trajectory of the components $m_{\alpha},\alpha=x,y,z$
for $\Xi/\epsilon_{\mathrm{F}}=0$ (left) and $\Xi/\epsilon_{\mathrm{F}}=0.05$
(right). As shown on the left, the $m_{z}$ component remains constant,
while the $m_{x,y}$ components undergo periodic oscillations, indicating
that the net magnetic moment is executing a precessional rotation
around the $z$-axis. In the presence of the splitting field $\Xi$
(shown on the right), the $z$ component of the magnetic moment of
the effective spin system exhibits oscillations characteristic of
nutational motion. Simultaneously, the $m_{x,y}$ components display
oscillatory motion (precession) with a period different from the case
where $\Xi=0$ (on the left). As previously explained, this is a result
of the anisotropy in the RKKY exchange coupling ($\Xi\neq0$).

As discussed in the introduction, numerous studies have examined magnetic
nutation using the macrospin approach (e.g., Refs. \cite{ciorneietal11prb,oliveetal12apl,DeEtAl_arxiv24}),
attributing the phenomenon to magnetic inertia, represented by an
additional time derivative, $d^{2}\bm{m}/dt^{2}$, in the LLG equation.
In our previous work \cite{basvenkac_prb18,AdamsEtAl_prb24_110},
we demonstrated through an atomistic approach that magnetization nutation
can be induced by surface and boundary effects. Similarly, the results
shown in Fig. \ref{fig:NumNut} here indicate that in magnetic
semiconductors, nutation arises from the anisotropy of the distance-dependent
long-range exchange coupling between localized spins. Thus, both nanomagnets
with surface effects and magnetic semiconductors with anisotropic
RKKY coupling exemplify cases where magnetic inertia---commonly thought
to induce nutation---is actually provided by (nonadiabatic) contributions
from local fluctuations and environmental degrees of freedom (see
similar discussions in Refs. \cite{kiktat_prb15,MakhlufEtAl_apl20,ThiNico_epjb21}).
This obviates the need for an additional time derivative in the magnetization's
evolution equation. Other examples include studies of nutation-like
excitations in systems of coupled vortices \cite{LeeEtAl_jap17}
and elongated nano-elements \cite{UrzOcam_j3m20}.

More generally, as discussed in Ref. \cite{basvenkac_prb18},
nutation arises whenever the effective field is time-dependent, as
illustrated in works 5--8 cited therein. Specifically, in the presence
of anisotropy, the effective field depends on the magnetic moment,
which is itself time-dependent. From a microscopic perspective, nutation
effects can be induced by fast degrees of freedom in the system, as
seen both in the present study on magnetic semiconductors and that in Ref.
\cite{basvenkac_prb18} on nanomagnets.
In the current work, nutation occurs in magnetic semiconductors due to
band splitting induced by an external magnetic field, implying that
the observed nutation is field-dependent. However, band splitting
could also be caused by internal dipolar fields in other systems.

A word is also in order regarding the effect of damping. For nanomagnets,
this was analyzed in Ref. \cite{basvenkac_prb18}
{[}see Fig. 6{]} within the atomistic approach. The impact of damping
on the time trajectories shown in Fig. \ref{fig:NumNut}
will be addressed in future work. More generally, in many-spin systems---such
as nanomagnets with isotropic exchange coupling (but with surface
anisotropy) and magnetic semiconductors with anisotropic RKKY exchange
coupling --- damping arises from spin-spin and spin-lattice interactions,
as well as multi-magnon processes. These mechanisms lead to the decay
of the excitation modes and the associated magnetization oscillations.
In the macroscopic approach, Lomonosov et al. \cite{LomonosovEtal_prb21}
discussed the inertial term ($d^{2}\bm{m}/dt^{2}$) which they included
with a coefficient $\tau=\beta/\alpha$, representing the relaxation
time of the fast degrees of freedom. They emphasize that while $\tau$
has a clear physical meaning, only $\beta$ is intrinsic to the material.
With such a convention, the inertial term is appropriately decoupled
from the Gilbert damping.

In the following section, we will employ the Green's function technique to derive the energy dispersion for the excitation modes associated with the precessional and nutational oscillations observed. Specifically, we will obtain analytical expressions for the corresponding frequencies as functions of the band-splitting field $\Xi$ and compare these results with the numerical solutions of Eqs. (\ref{eq:LLE}).

\subsection{\label{sec:precnut-freq}Precession and nutation frequencies: an analytical study, comparison with the numerical solution}

Combining the initial Hamiltonian (\ref{eq:Ham-LocalizedSpins}) for
the localized spins with the contribution induced by their coupling
to the conduction electrons, namely the contribution in Eq. (\ref{eq:XXZHam}),
we obtain the total Hamiltonian for the localized spins in the magnetic
semiconductor 
\begin{equation}
\mathcal{H}_{\mathrm{FMS}}=-\sum_{i,j}\tilde{\mathcal{J}}_{\perp}\left(\bm{r}_{ij}\right)\left(S_{i}^{x}S_{j}^{x}+S_{i}^{y}S_{j}^{y}\right)-\sum_{i,j}\mathcal{\tilde{J}}_{\parallel}\left(\bm{r}_{ij}\right)S_{i}^{z}S_{j}^{z}-g_{S}\mu_{\mathrm{B}}H_{0}\cdot\sum_{i}S_{i}^{z}\label{eq:H-XXZ}
\end{equation}
with $\tilde{\mathcal{J}}_{\perp}\left(\bm{r}_{ij}\right)=J+\mathcal{J}_{\perp}\left(\bm{r}_{ij}\right)$
and $\mathcal{\tilde{J}}_{\parallel}\left(\bm{r}_{ij}\right)=J+\mathcal{J}_{\parallel}\left(\bm{r}_{ij}\right)$.

Defining the retarded Green's function 
\begin{equation}
G^{(r)}(i-j,t)=-i\theta(t)\left\langle \left[S_{i}^{-}(t),S_{j}^{+}(0)\right]\right\rangle \label{FM2}
\end{equation}
and using its equation of motion 
\begin{eqnarray*}
i\frac{dG^{(r)}(i-j,t)}{dt} & = & \delta(t)\left\langle \left[S_{i}^{-}(0),S_{j}^{+}(0)\right]\right\rangle +\theta(t)\left\langle \left[\frac{dS_{i}^{-}(t)}{dt},S_{j}^{+}(0)\right]\right\rangle 
\end{eqnarray*}
together with the \emph{SU(2)} spin algebra $\left[S_{i}^{z},S_{j}^{\mu}\right]=\mu S_{i}^{\mu}\delta_{ij},\left[S_{i}^{+},S_{j}^{-}\right]=2S_{i}^{z}\delta_{ij},\mu=\pm$,
we obtain the spin-wave dispersion relation $\omega\left(\bm{k}\right)$
as the solution of the equation 
\begin{equation}
\hbar\omega\left(\bm{k},\Xi\right)=g_{S}\mu_{\mathrm{B}}H_{0}+2zJ\left\langle S^{z}\right\rangle \left(1-\gamma_{\bm{k}}\right)+2\left\langle S^{z}\right\rangle \left[\mathcal{J}_{\parallel}\left(\mathbf{0},\omega\left(\bm{k}\right),\Xi\right)-\mathcal{J}_{\perp}\left(\mathbf{k},\omega\left(\bm{k}\right),\Xi\right)\right].\label{eq:SW-energy}
\end{equation}

Here, $\left\langle S^{z}\right\rangle $ is the average magnetization.
We have also used Bogoliubov and Tyablikov\citep{bogtya59} decoupling
approximation 
\begin{equation}
-i\theta(t)\left\langle \left[(S_{k}^{z}S_{i}^{-})(t),S_{j}^{+}(0)\right]\right\rangle =\left\langle S^{Z}\right\rangle \;G_{ij}^{(r)}(t)\label{eq:FM4-2}
\end{equation}
with the assumption $\left\langle S_{i}^{z}\right\rangle =\left\langle S^{z}\right\rangle $.

In the absence of the \emph{sd} coupling, Eq. (\ref{eq:SW-energy})
becomes 
\begin{align*}
\omega\left(\bm{k},\xi\right) & =g_{S}\mu_{\mathrm{B}}H_{0}+2zJ\left\langle S^{z}\right\rangle \left(1-\gamma_{\bm{k}}\right).
\end{align*}
which is the usual dispersion law of a ferromagnet with isotropic
exchange coupling.

In the static limit, \emph{i.e.} setting $\omega=0$ on the right
hand side of (\ref{eq:SW-energy}), we have (in the approximation
$\Xi\ll\epsilon_{\mathrm{F}}$) 
\begin{align}
\hbar\omega\left(\bm{k},\Xi\right) & \simeq g_{S}\mu_{\mathrm{B}}H_{0}+2zJ\left\langle S^{z}\right\rangle \left(1-\gamma_{\bm{k}}\right)+2\left\langle S^{z}\right\rangle \left[\mathcal{J}_{\parallel}\left(\mathbf{0},0,\Xi\right)-\mathcal{J}_{\perp}\left(\mathbf{k},0,\Xi\right)\right].\label{eq:SWDL}
\end{align}

The full analysis of the excitation modes and their experimental characterization
will be presented in a future work. Using a similar technique of averaging
out the conduction electron degrees of freedom along with the Green's
function and self-energy approach, the effect of band occupation on
the spin-wave dispersion and thereby on the critical temperature was
studied in the framework of Kondo-lattice model\citep{SantosNolting_prb02},
see also \citep{KonigEtAl_prl00,KonigEtAl_prb01}.

Here, we focus on the first effects of the \emph{sd} coupling on the
precession frequency (uniform mode) and the low-lying modes, and how
they induce nutation of the magnetization of the effective spin system.

Let us denote by $\hbar\Delta\omega_{\mathrm{sd}}\left(\bm{k},\Xi\right)$
the contribution to $\hbar\omega$ stemming from the interaction between
the conduction electrons and localized spins, namely {[}see Eq. (\ref{eq:SW-energy}){]}
\begin{equation}
\hbar\Delta\omega_{\mathrm{sd}}\left(\bm{k},\Xi\right)=2\left\langle S^{z}\right\rangle \left[\mathcal{J}_{\parallel}\left(\mathbf{0},0,\Xi\right)-\mathcal{J}_{\perp}\left(\mathbf{k},0,\Xi\right)\right].\label{eq:SWDL_sd}
\end{equation}

Using the fact that $\mathcal{F}\left(0\right)=1$, Eq. (\ref{eq:Jparal_mumu})
yields ($\tilde{\Xi}=\Xi/\epsilon_{\mathrm{F}}$)
\[
\mathcal{J}_{\parallel}\left(\mathbf{k},0,\Xi\right)=-\frac{\lambda^{2}}{16}\left(\rho_{\mathrm{F}}^{\uparrow}+\rho_{\mathrm{F}}^{\downarrow}\right)=\frac{\lambda^{2}\rho_{\mathrm{F}}}{16}\left(\sqrt{1+\tilde{\Xi}}+\sqrt{1-\tilde{\Xi}}\right)
\]
and for small $\Xi$, $\mathcal{J}_{\parallel}\left(\mathbf{k},0,\Xi\right)\simeq\lambda^{2}\rho_{\mathrm{F}}/8$.

For the coupling $\mathcal{J}_{\perp}\left(\mathbf{k},0,\Xi\right)$,
we may proceed by an expansion in the long-wavelength limit ($k\rightarrow0$)
using the following expansion ($a\neq0$) 
\begin{equation}
\left(1+\frac{a}{t^{2}}\right)\mathcal{F}\left(\kappa\left(1+\frac{a}{\kappa^{2}}\right)\right)\simeq\frac{1}{3a}-\frac{\left(5a-1\right)}{15a^{3}}\kappa^{2},\label{eq:LindhardExpanded}
\end{equation}
with $a=\tilde{\Xi}/2$ and $\kappa=k/2p_{\mathrm{F}}^{\uparrow}$.

Therefore, for the uniform or precession mode ($\bm{k}=\bm{0}$) we
obtain 
\begin{equation}
\hbar\Delta\omega_{\mathrm{prec}}\left(\bm{0},\Xi\right)=2\left\langle S^{z}\right\rangle \frac{\lambda^{2}\rho_{\mathrm{F}}}{8}\left[\frac{\left(1+\tilde{\Xi}\right)^{3/2}-\left(1-\tilde{\Xi}\right)^{3/2}}{3\tilde{\Xi}}-\frac{1}{2}\left(\sqrt{1+\tilde{\Xi}}+\sqrt{1-\tilde{\Xi}}\right)\right].\label{eq:SWDL_sd_v2}
\end{equation}

\begin{figure}[H]
\begin{centering}
\includegraphics[scale=0.6]{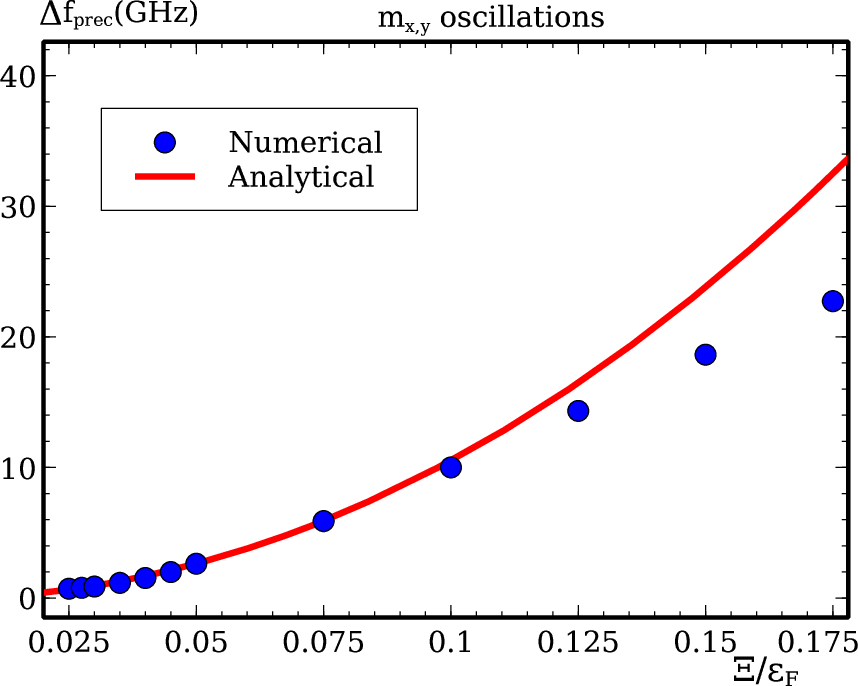}
\par\end{centering}
\caption{\label{fig:DeltaOmega}Correction to the precession frequency ($\Delta f_{\mathrm{prec}}=\Delta\omega_{\mathrm{prec}}/2\pi$)
due to the \emph{sd }coupling as given (in red) by the quantity in
square brackets in Eq. (\ref{eq:SWDL_sd_v2}) and (in blue symbols)
by the numerical simulation.}
\end{figure}

The term between square brackets is an increasing function of the
splitting field $\Xi$. This implies that the precession frequency
($f=\omega/2\pi$) of the localized spins increases with $\tilde{\Xi}$.
This result is confirmed by the numerical simulations of Section \ref{subsec:Exchange-coupling-inFS},
see comparison in Fig. \ref{fig:DeltaOmega}. For small $\tilde{\Xi}$,
the precession frequency of the localized spins is given by 
\begin{equation}
\hbar\omega\left(\bm{0},\Xi\right)\simeq g_{S}\mu_{\mathrm{B}}H_{0}+\left\langle S^{z}\right\rangle \frac{\lambda^{2}\rho_{\mathrm{F}}v_{0}}{48}\tilde{\Xi}^{2}.\label{eq:PreceFreq_smallXi}
\end{equation}

In a typical magnetic semiconductor for which $\lambda^{2}\rho_{\mathrm{F}}v_{0}/8\simeq10\,\mathrm{meV}$,
the last term leads to a frequency on the order of $10\,\mathrm{GHz}$
for $\tilde{\Xi}=0.1$.

For the nonuniform mode ($\bm{k}\neq\bm{0}$), we obtain in the long-wavelength
limit 
\begin{align}
\hbar\Delta\omega_{\mathrm{nut}}\left(k\rightarrow0,\Xi\right) & \equiv\hbar\Delta\omega_{\mathrm{sd}}\left(\bm{k},\xi\right)-\hbar\Delta\omega_{\mathrm{prec}}\left(\bm{0},\Xi\right)\label{eq:SWDL_sd_nut}\\
 & =2\left\langle S^{z}\right\rangle \frac{\lambda^{2}\rho_{\mathrm{F}}}{8}\frac{16}{5}\left[\frac{\left(1-\tilde{\Xi}\right)^{3/2}\left(1+\frac{3}{2}\tilde{\Xi}\right)-\left(1-\frac{3}{2}\tilde{\Xi}\right)\left(1+\tilde{\Xi}\right)^{3/2}}{\tilde{\Xi}^{3}}\right]k^{2}.\nonumber 
\end{align}

For the same parameters as before, this contribution is in the THz
range. Accordingly, as discussed in Ref. \cite{MyersEtAl_prb05} and
references therein, in nonmagnetic conventional semiconductor hetero-structures,
the variation of the spin-orbit coupling allows for an electrical
control of coherent electron-spin dynamics involving energies on the
order of the $\mu$eV, leading to spin precession frequencies in the GHz
range. In comparison, the \emph{sd} exchange interaction in magnetic
semiconductors induces large spin splittings in the conduction band
states with energies on the order of $10\lyxmathsym{\textendash}100$
meV, or frequencies in the THz regime \citep{AwschalomSmarth02}.

The plot in Fig. \ref{fig:DeltaOmega} shows that the precession frequency
is also a function of $\tilde{\Xi}$. More precisely, we plot (in
red) the correction to the precession frequency, due to the \emph{sd
}coupling, given by the quantity in square brackets in Eq. (\ref{eq:SWDL_sd_v2})
for the uniform mode ($k=0$). The blue symbols represent the result
rendered by the numerical simulation of the effective spin system
with space-dependent effective exchange couplings, presented in Section
\ref{subsec:Magnetic-nutation_LLE}. The good match between the analytical
and numerical results (for small $\Xi$) shows that the effective
XXZ spin model confirms the predictions of our general formalism in
the presence of a magnetic field.

\begin{figure}[H]
\begin{centering}
\includegraphics[scale=0.7]{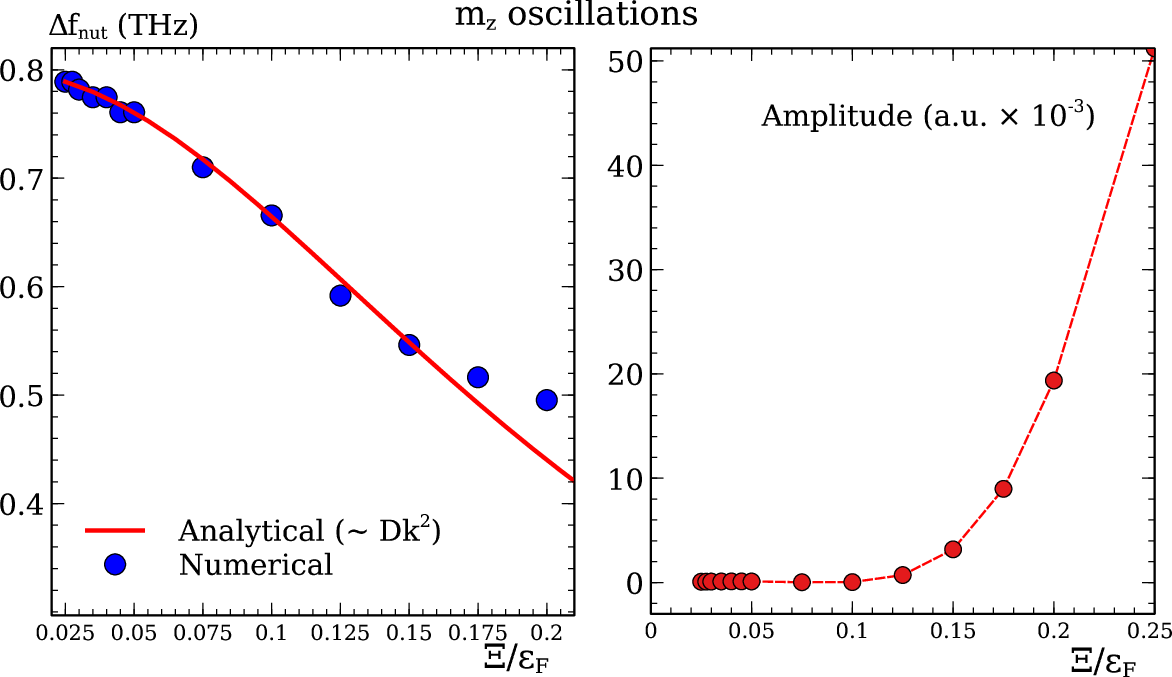}
\par\end{centering}
\caption{\label{fig:FreqAmp_nut}Frequency ($\Delta f_{\mathrm{nut}}=\Delta\omega_{\mathrm{nut}}/2\pi$)
and amplitude of the $m_{z}$ oscillations against the (effective)
splitting parameter $\Xi/\epsilon_{F}$.}
\end{figure}

For the nonuniform mode ($k\neq0$), we plot in Fig. \ref{fig:FreqAmp_nut}
(left) the frequency $\Delta f_{\mathrm{nut}}\equiv\Delta\omega_{\mathrm{nut}}/2\pi$,
where $\Delta\omega_{\mathrm{nut}}$ is given by Eq. (\ref{eq:SWDL_sd_nut})
for $ak\simeq0.0468$ (in red line) and in symbols the result of the
numerical simulations {[}Section \ref{subsec:Magnetic-nutation_LLE}{]}.
On the right, we plot the amplitude of the corresponding nutational
mode as rendered by the numerical study of the dynamics of the net
magnetic moment. Here again, the good agreement between the analytical
expression and the numerical result in Fig. \ref{fig:FreqAmp_nut}
(left), shows that the analytical developments using the effective
spin model provide a good approximation for the solution of the Landau-Lifshitz
equation, at least for materials with $\Xi$ small with respect to
$\epsilon_{\mathrm{F}}$. On the other hand, the frequency plot shows
that the RKKY result deviates from that rendered by the $k^{2}$ expansion
at a relatively high values of the splitting parameter $\Xi$. A similar
result was obtained in Ref. \cite{KonigEtAl_prl00} which shows, for
a given $\Xi$, such a deviation of the long-wavelength spin-wave
branch from that of the mode induced by the sd coupling. On the other
hand, the plot on the right clearly shows that the amplitude of the
nutation mode increases with $\Xi$. Note that the derivation of an
analytical expression for the amplitude of the nutation mode is somewhat
more involved and will be postponed to a subsequent work (\emph{e.g.}
using a similar approach to that of Ref. \cite{suhl57jcps}). 

\section{Conclusion}

In this study, we have demonstrated, for the first time, that magnetization
in magnetic semiconductors exhibits nutational motion when subjected
to an external magnetic field. This nutation arises from the band
splitting of conduction electrons which leads to anisotropic, distance-dependent
exchange coupling between localized spins. Using a comprehensive theoretical
framework, we modeled the system as consisting of conduction electrons
and a ferromagnetic array of localized spins, coupled through the
Vonsovskii (sd) interaction. 

By applying the Feynman-Schwinger formalism, we derived an effective
spin Hamiltonian after integrating out the charge degrees of freedom,
which allowed us to focus on the spin dynamics. Our work extends the
RKKY formalism to include external electrical and magnetic disturbances,
providing an effective spin Hamiltonian with explicit expressions
for the long-range exchange couplings in both Fourier and direct spaces.
Then, through numerical solution of the coupled Landau-Lifshitz equations
for the localized spins, we analyzed the dynamics of the net magnetic
moment, identifying both nutational and precessional modes. We found
that the nutation frequency decreases while its amplitude increases
with the strength of the band-splitting field. In contrast, the precession
frequency rises, reaching variations in the tens of GHz range, while
low-lying nonuniform modes appear in the THz regime, consistent with
earlier studies in other materials. Additionally, we derived analytical
expressions for precession and nutation frequencies, confirming their
consistency with numerical results.

Our findings show that nutation in magnetic semiconductors can occur without requiring the second time derivative in the magnetization
equation of motion. Instead, here nutation is triggered by nonadiabatic contributions from environmental degrees of freedom (charge and local
spin fluctuations) similar to the behavior seen in nanomagnets with surface anisotropy.

This work establishes a solid basis for advancing research on nutational dynamics in magnetic semiconductors, with exciting applications in spintronics and optoelectronic control of magnetic properties. Future studies will expand on these findings by examining the influence of magneto-crystalline anisotropy, damping, crystal structure, doping, and multi-band models to uncover more intricate nutational spectra and enhance our understanding of these dynamics in practical devices. Such advancements are crucial for understanding and manipulating spin behavior on pico- and femtosecond timescales, with significant implications for ultra-fast, energy-efficient data processing and storage technologies.

\acknowledgments The author thanks A. Michels and U. Nowak for reading the manuscript.


\newpage{}
\appendix

\section{\label{sec:Effective-action}Effective action for localized spins:
a path-integral approach}

In this appendix, we present the details of our formalism leading
to the final expression in Eq. (\ref{eq:Aeff-noPP}) for the action
of the magnetic subsystem of localized magnetic ions.

In this approach, instead of computing the grand-canonical partition
function and its derivatives with respect to $\beta=1/k_{{\rm B}}T$
to infer the various thermodynamic functions, one represents the grand-canonical
partition function as a path integral over Grassmann coherent states
\citep{negorl98}. This amounts to replacing, at finite temperature,
the creation and annihilation operators, $C_{{\bf k}\alpha}^{\dagger}\left(t\right),C_{{\bf k}\alpha}\left(t\right)$,
in the Heisenberg representation, by the imaginary-time dependent
Grassmann fields $\psi_{{\bf k}\alpha}^{*}\left(\tau=it\right),\psi_{{\bf k}\alpha}\left(\tau=it\right)$,
respectively. Note that whereas $C_{{\bf k}\alpha}^{\dagger}$ is
the adjoint of $C_{{\bf k}\alpha}$, the two Grassmann fields $\psi_{{\bf k}\alpha}^{*}$
and $\psi_{{\bf k}\alpha}$ are independent of each other.

\subsection{Euclidean action}

Using these new variables, the grand-canonical partition function
reads {[}see \emph{e.g. }Refs. \citep{ryder96,negorl98,MudryWS2014}{]}
\begin{equation}
Z={\rm Tr}\,e^{-\beta\mathcal{H}}=\int{\cal D}\left[\bm{S}\right]{\cal D}\left[\psi^{*}\right]{\cal D}\left[\psi\right]\,e^{-\mathcal{A}_{{\rm E}}}\label{eq:GCPF}
\end{equation}
where $\mathcal{A}_{{\rm E}}$ is the Euclidean action (of the general
form $\mathcal{A}_{{\rm E}}=\int_{0}^{\beta}d\tau\,\left[\eta^{*}\partial_{\tau}\eta+\mathcal{H}\left(\eta^{*},\eta\right)\right]$)
\begin{equation}
\mathcal{A}_{{\rm E}}=\intop_{0}^{\beta}d\tau\,\sum_{\alpha,\beta}\sum_{{\bf k},{\bf p}}\psi_{{\bf k}\alpha}^{*}\left(\tau\right)\Lambda_{\alpha\beta}\left({\bf k},{\bf p};\tau\right)\psi_{{\bf p}\beta}\left(\tau\right)\label{eq:EuclideanAction}
\end{equation}
with the bi-linear form

\begin{align}
\Lambda_{\alpha\beta}\left({\bf k},{\bf p};\tau\right) & =\Big[\delta_{\alpha\beta}\partial_{\tau}+E_{{\bf k}}\delta_{\alpha\beta}-\left(\bm{\xi}\cdot{\bf \bm{\sigma}}_{\alpha\beta}\right)\Big]\delta_{\bm{kp}}+\frac{1}{V}\Upsilon\left({\bf k}-{\bf p},\tau\right).\label{eq:BilinearForm}
\end{align}

Note that the Grassmann integration variables obey the anti-periodic
boundary conditions in imaginary time\citep{FetterWalecka71,MudryWS2014}
\begin{equation}
\psi_{{\bf k}\alpha}^{*}\left(\tau+\beta\right)=-\psi_{{\bf k}\alpha}^{*}\left(\tau\right),\quad\psi_{{\bf k}\alpha}\left(\tau+\beta\right)=-\psi_{{\bf k}\alpha}\left(\tau\right).\label{eq:APBC}
\end{equation}

Next, we discuss the integration measure ${\cal D}\left[\bm{S}\right]{\cal D}\left[\psi^{*}\right]{\cal D}\left[\psi\right]$
in Eq. (\ref{eq:GCPF}). For the (local) spin variables $S_{\alpha}$,
we have ($S_{\alpha}=Ss_{\alpha}$ ) 
\[
{\cal D}\left[\bm{S}\right]=\prod_{\alpha=1}^{\mathcal{N}}ds_{\alpha}=\prod_{\alpha=1}^{\mathcal{N}}\sin\theta_{\alpha}d\varphi_{\alpha}.
\]

For the Grassmann variables, we first write 
\begin{align}
\psi_{{\bf k}\alpha}^{*}\left(\tau\right) & =\frac{1}{\sqrt{\beta}}\sum_{n\in\mathbb{Z}}\psi_{{\bf k},\omega_{n},\alpha}^{*}e^{i\omega_{n}\tau},\label{eq:FT-psis}\\
\psi_{{\bf k}\alpha}\left(\tau\right) & =\frac{1}{\sqrt{\beta}}\sum_{n\in\mathbb{Z}}\psi_{{\bf k},\omega_{n},\alpha}e^{-i\omega_{n}\tau},\nonumber 
\end{align}
where $\omega_{n}$ are the Matsubara frequencies given by 
\begin{equation}
\omega_{n}=\frac{\pi}{\beta}\left(2n+1\right),\quad n\in\mathbb{Z}\label{eq:MatsubaraFreq}
\end{equation}
with

\begin{equation}
\frac{1}{\beta}\intop_{0}^{\beta}d\tau\,e^{i\left(\omega_{n}-\omega_{m}\right)\tau}=\delta_{mn}.\label{eq:Integration-MatsubaraFreq}
\end{equation}

Then, we have the measure for the Grassmann variables 
\[
{\cal D}\left[\psi^{*}\right]{\cal D}\left[\psi\right]=\prod_{\bm{k},n,\alpha}d\psi_{{\bf k},\omega_{n},\alpha}^{*}d\psi_{{\bf k},\omega_{n},\alpha}.
\]

For the bosonic fields, we use the Fourier transforms

\begin{align*}
\Phi_{{\rm pl}}\left({\bf q},\tau\right) & =\frac{1}{\beta}\sum_{l\in\mathbb{Z}}e^{i\Omega_{l}\tau}\Phi_{{\rm pl}}\left({\bf q},\Omega_{l}\right),\\
{\bf S}\left({\bf q},\tau\right) & =\frac{1}{\beta}\sum_{l\in\mathbb{Z}}e^{i\varpi_{l}\tau}{\bf S}\left({\bf q},\varpi_{l}\right),
\end{align*}
with $\Omega_{l}=2\pi l/\beta$ and $\varpi_{l}=2\pi l/\beta$ being
bosonic Matsubara frequencies.

Therefore, Eq. (\ref{eq:EuclideanAction}) becomes
\begin{widetext}
\begin{align}
\mathcal{A}_{{\rm E}} & =\sum_{\alpha,\beta}\sum_{{\bf k},{\bf p}}\sum_{n,m\in\mathbb{Z}}\psi_{{\bf k},\omega_{n},\alpha}^{*}\left(-i\omega_{n}\right)\delta_{\alpha\beta}\delta_{\bm{kp}}\delta_{mn}\psi_{{\bf p},\omega_{m},\beta}\label{eq:EclideanAction-psis}\\
 & +\sum_{\alpha,\beta}\sum_{{\bf k},{\bf p}}\sum_{n,m\in\mathbb{Z}}\psi_{{\bf k},\omega_{n},\alpha}^{*}\Big[E_{{\bf k}}\delta_{\alpha\beta}-\left(\bm{\xi}\cdot{\bf \bm{\sigma}}_{\alpha\beta}\right)\Big]\delta_{\bm{kp}}\delta_{mn}\psi_{{\bf p},\omega_{m},\beta}\nonumber \\
 & +\sum_{\alpha,\beta}\sum_{{\bf k},{\bf p}}\sum_{n,m\in\mathbb{Z}}\psi_{{\bf k},\omega_{n},\alpha}^{*}\frac{1}{\beta V}\left[\Phi_{{\rm pl}}\left({\bf k}-{\bf p},\Omega_{m-n}\right)\delta_{\alpha\beta}-\frac{\lambda}{2}{\bf S}\left({\bf k}-{\bf p},\varpi_{m-n}\right)\cdot\bm{\sigma}_{\alpha\beta}\right]\delta_{mn}\psi_{{\bf p},\omega_{m},\beta}\nonumber 
\end{align}
\end{widetext}

where $\varpi_{m-n}=\varpi_{m}-\varpi_{n}=2\pi\left(m-n\right)/\beta$,
similarly for $\Omega_{m-n}$.

To simplify the notation, we will use the short hand notation $k=\left(\mathbf{k},n\right),p=\left(\mathbf{p},m\right),q=\left(\mathbf{q},l\right)$
, etc, and rewrite the action as follows: 
\begin{align*}
\mathcal{S}_{{\rm E}} & =\sum_{k,p}\sum_{\alpha,\beta}\psi_{k,\alpha}^{*}\Lambda_{\alpha\beta}^{nm}\left(k,p\right)\psi_{p,\beta}
\end{align*}
with 
\begin{align}
\Lambda_{\alpha\beta}^{nm}\left(k,p\right) & =-\Big[\left(i\omega_{n}-E_{{\bf k}}\right)\delta_{\alpha\beta}+\left(\bm{\xi}\cdot{\bf \bm{\sigma}}_{\alpha\beta}\right)\Big]\delta_{kp}+\frac{1}{\beta V}\left[\Phi_{{\rm pl}}\left(k-p\right)\delta_{\alpha\beta}-\frac{\lambda}{2}{\bf S}\left(k-p\right)\cdot\bm{\sigma}_{\alpha\beta}\right]\label{eq:MatrixLambda}
\end{align}
where $\delta_{kp}\equiv\delta_{\bm{kp}}\delta_{mn}$ and $\left(k-p\right)\rightarrow\left({\bf k}-{\bf p},\varpi_{m-n}\right)$,
remembering that for the spin variables we have the frequencies $\varpi_{m}$
while for the external charge disturbance, these are $\Omega_{m}$.

Summing up, the previous developments, we write

\begin{align*}
Z & =\int{\cal D}\left[\bm{S}\right]{\cal D}\left[\psi^{*}\right]{\cal D}\left[\psi\right]\,e^{-\mathcal{S}_{{\rm E}}}=\int{\cal D}\left[\bm{S}\right]\times\prod_{\bm{k},n,\alpha}d\psi_{{\bf k},\omega_{n},\alpha}^{*}d\psi_{{\bf k},\omega_{n},\alpha}\,e^{-\mathcal{A}_{{\rm E}}}\\
 & =\int{\cal D}\left[\bm{S}\right]\times\prod_{\bm{k},n,\alpha}d\psi_{{\bf k},\omega_{n},\alpha}^{*}d\psi_{{\bf k},\omega_{n},\alpha}\exp\left\{ -\sum_{k,p}\sum_{\alpha,\beta}\psi_{k,\alpha}^{*}\Lambda_{\alpha\beta}^{nm}\left({\bf k},{\bf p}\right)\psi_{p,\beta}\right\} .
\end{align*}

Then, after integration over the Grassmann variables using $\int dc^{\ast}dc\,e^{-\int c^{\ast}Mc}=\det M$,
we obtain 
\begin{align*}
Z & =\int{\cal D}\left[\bm{S}\right]\times\det\Lambda=\int{\cal D}\left[\bm{S}\right]\times e^{\ln\left(\det\Lambda\right)}\equiv\int{\cal D}\left[\bm{S}\right]\times e^{-\mathcal{A}_{{\rm eff}}}
\end{align*}
where the effective $\mathcal{A}_{{\rm eff}}$ action is given by
$\mathcal{A}_{{\rm eff}}=-\ln\left(\det\Lambda\right)=-\Tr\ln\left(\Lambda\right)$;
the trace includes a trace on both the conduction electron variables
and the spin variables.

\subsection{Expanding in \emph{sd} coupling}

In the one-band model, the conduction band $E_{\bm{k}}=z\eta\gamma_{\bm{k}},\,\gamma_{\bm{k}}=\sum_{\bm{a}}e^{i\bm{k}\cdot\bm{a}}/z$,
is very large as compared to the \emph{sd }exchange coupling $\lambda S$
{[}see Eq. (\ref{eq:Ham-sdCoupling}){]}. In FMS, such as europium
chalcogenides (EuS), the band width is of $3-5$ eV and the \emph{sd}
coupling is circa $0.5$ eV {[}see the review \cite{MaugerGoddart_PR86}{]}.
A similar situation applies in the case of DMS {[}see textbook \cite{GajKos_spr11}
and Ref. \cite{BeaDan_prb10}{]}. As such, we may perform an expansion
with respect to the coupling constant $\lambda$, see also discussion
in Ref. \cite{BabcencoCottam_JPhysCSoldStatPhys.14.5347}. For this
purpose, we rewrite the tensor in Eq. (\ref{eq:MatrixLambda}) as
$\Lambda=\mathcal{F}+\Pi$\emph{, i.e.} splitting $\Lambda$ into
two parts $\mathcal{F}$ and $\Pi$ whose components are given by
\begin{align}
\mathcal{F}_{\alpha\beta}^{nm}\left(k,p\right) & =-\Big[\left(i\omega_{n}-E_{{\bf k}}\right)\delta_{\alpha\beta}+\left(\bm{\xi}\cdot{\bf \bm{\sigma}}_{\alpha\beta}\right)\Big]\delta_{kp}+\frac{1}{\beta V}\Phi_{{\rm pl}}\left(k-p\right)\delta_{\alpha\beta},\nonumber \\
\Pi_{\alpha\beta}^{nm}\left(k,p\right) & =-\frac{\lambda}{2\beta V}{\bf S}\left(k-p\right)\cdot\bm{\sigma}_{\alpha\beta}.\label{eq:FXi}
\end{align}

$\mathcal{F}$ represents the free contribution and $\Xi$ that of
the $\lambda$ interaction, considered here as a perturbation. Then,
dropping the first term which does not contain spin operators, we
write $\mathcal{A}_{{\rm eff}}=-{\rm Tr}\log\Lambda=\mathcal{A}_{{\rm eff}}^{\left(1\right)}+\mathcal{A}_{{\rm eff}}^{\left(2\right)}$,
where the $1^{\mathrm{st}}$ and $2^{\mathrm{nd}}$ order contributions
to the effective action are given by 
\begin{align}
\mathcal{A}_{{\rm eff}}^{\left(1\right)} & =-{\rm Tr}\left[\mathcal{F}^{-1}\Pi\right],\nonumber \\
\mathcal{A}_{{\rm eff}}^{\left(2\right)} & =\frac{1}{2}{\rm Tr}\left[\left(\mathcal{F}^{-1}\Pi\right)^{\dagger}\left(\mathcal{F}^{-1}\Pi\right)\right].\label{eq:1st2nd-order-contribs}
\end{align}

The next task consists in computing the trace appearing above which
involves summing over the spin degrees of freedom, the wave vectors
and frequencies. We will first compute the trace over the (electron)
spin variables using the well known properties of the spin-1/2 algebra
of the Pauli matrices ${\bf \sigma}_{i},i=1,2,3$. For later reference,
we introduce a few new operators. First, we rewrite the operator $\mathcal{F}$
appearing in (\ref{eq:FXi}), which represents the contribution in
the absence of the \emph{sd }coupling, as

\begin{align}
\mathcal{F}_{\alpha\beta}^{nm}\left({\bf k},{\bf p}\right) & =F^{nm}\left({\bf k},{\bf p}\right)\delta_{\alpha\beta}-\delta_{\bm{kp}}\delta^{nm}\left(\bm{\xi}\cdot{\bf \bm{\sigma}}_{\alpha\beta}\right)\label{eq:FXI_F}
\end{align}
where 
\begin{align}
F^{nm}\left({\bf k},{\bf p}\right) & =F_{0}^{nm}\left({\bf k},{\bf p}\right)+\frac{1}{\beta V}\Phi_{{\rm pl}}\left({\bf k}-{\bf p},\Omega_{m-n}\right)\label{eq:F-noField}
\end{align}
is the contribution in the absence of the effective magnetic field
(and is thereby independent of spin variables) and

\begin{align}
F_{0}^{nm}\left({\bf k},{\bf p}\right) & =-\left(i\omega_{n}-E_{{\bf k}}\right)\delta^{nm}\delta_{\bm{kp}}\label{eq:FreeFreeF}
\end{align}
is the contribution in the absence of any kind of interactions. Note
that in tensor form, we may write $\mathcal{F}=F\otimes\mathds{1}_{{\rm s}}-\mathds{1}_{{\rm p}}\otimes\left(\bm{\xi}\cdot{\bf \bm{\sigma}}\right)$,
where $\mathds{1}_{{\rm s}},\mathds{1}_{{\rm p}}$ are the identity
matrices in the space of spin variables and wave vectors, respectively.

To compute the inverse of $\mathcal{F}$, needed in Eq. (\ref{eq:1st2nd-order-contribs}),
we note that, with regard to spin variables, $\mathcal{F}$ is a linear
combination of the unit matrix $\mathds{1}_{{\rm s}}$ and the three
Pauli matrices ${\bf \bm{\sigma}}$, forming a complete set of the
space of $2\times2$ matrices. Hence, we may seek $\mathcal{F}^{-1}$
in the form $\mathcal{F}^{-1}=M\mathds{1}_{{\rm s}}+\bm{\phi}\cdot{\bf \bm{\sigma}}$,
where $M$ is an $\mathcal{N}\times\mathcal{N}$ matrix in phase space
and $\bm{\phi}$ is a result of an $\mathcal{N}\times\mathcal{N}$
matrix acting on a vector. Then, using ${\rm Tr}\sigma^{i}=0$ and
$\left(\sigma^{i}\right)^{2}=\mathds{1}_{{\rm s}}$, we obtain the
coefficients $M$ and $\phi$. Doing so, we find that the inverse
of $\mathcal{F}$ is given by 
\begin{align}
\mathcal{F}^{-1} & =-\left(\xi^{2}\mathds{1}_{{\rm p}}-F^{2}\right)^{-1}\left[F\mathds{1}_{{\rm s}}+\left(\bm{\xi}\cdot{\bf \bm{\sigma}}\right)\right]\nonumber \\
 & \equiv-\mathcal{D}\left[F\mathds{1}_{{\rm s}}+\left(\bm{\xi}\cdot{\bf \bm{\sigma}}\right)\right]\label{eq:CalFInverse}
\end{align}
where we have introduced the tensor 
\begin{equation}
\mathcal{D}\equiv\left(\xi^{2}\mathds{1}_{{\rm p}}-F^{2}\right)^{-1}=\left(\xi\mathds{1}_{{\rm p}}-F\right)^{-1}\left(\xi\mathds{1}_{{\rm p}}+F\right)^{-1}\label{eq:TensorD}
\end{equation}
which is also diagonal in spin variables. We then obtain the result

\begin{align}
\mathcal{F}^{-1}\Xi & =\frac{\lambda}{2\beta V}\mathcal{D}\left(\bm{\xi}\cdot{\bf \bm{\sigma}}+F\mathds{1}_{{\rm s}}\right)\times\left[{\bf S}\left(k-p\right)\cdot\bm{\sigma}\right].\label{eq:FinvXi}
\end{align}

Finally, we define the Green's function 
\begin{align}
\mathcal{G} & \equiv-\mathcal{F}=\mathcal{D}\left(F\mathds{1}_{{\rm s}}+\bm{\xi}\cdot{\bf \bm{\sigma}}\right)\nonumber \\
 & =\left(\xi\mathds{1}_{{\rm p}}-F\right)^{-1}\left(\xi\mathds{1}_{{\rm p}}+F\right)^{-1}\left(F\mathds{1}_{{\rm s}}+\bm{\xi}\cdot{\bf \bm{\sigma}}\right).\label{eq:FreePropagator}
\end{align}

It is easy to see that when there is no external disturbance ($\Phi_{{\rm pl}}\equiv0$,
$\xi=0$), $\mathcal{G}$ becomes the ``free propagator'' $\mathcal{G}_{0}=-F^{-1}$
with the components 
\begin{equation}
\boxed{\left(\mathcal{G}_{0}\right)_{\alpha\beta}^{nm}\left({\bf k},{\bf p}\right)=\frac{1}{i\omega_{n}-E_{\bm{k}}}{\bf \delta}_{\alpha\beta}\delta_{\bm{kp}}\delta_{nm}}.\label{eq:Free-e-Propagator}
\end{equation}

Next, to obtain the $1^{\mathrm{st}}$ and $2^{\mathrm{nd}}$ order
contributions, we compute the traces in Eq. (\ref{eq:1st2nd-order-contribs})
using the following identities (which are well known properties of
the spin-1/2 algebra) 
\begin{align}
\left({\bf \bm{\sigma}}\cdot\bm{a}\right)\left({\bf \bm{\sigma}}\cdot\bm{b}\right) & =\left({\bf \bm{a}}\cdot\bm{b}\right)\mathds{1}_{\mathrm{s}}+i{\bf \bm{\sigma}}\cdot\left({\bf \bm{a}}\times\bm{b}\right).\label{eq:sigmaProps}\\
{\rm Tr}\sigma^{i} & =0,\nonumber \\
{\rm Tr}\left({\bf \sigma}^{i}{\bf \sigma}^{j}\right) & =2\delta^{ij},\,i,j=1,2,3,\nonumber \\
{\rm Tr}\left({\bf \sigma}^{i}{\bf \sigma}^{j}\sigma^{k}\right) & =2i\epsilon{}^{ijk},\nonumber \\
{\rm Tr}\left({\bf \sigma}^{i}{\bf \sigma}^{j}\sigma^{k}\sigma^{l}\right) & =2\left(\delta^{ij}\delta^{kl}-\delta^{ik}\delta^{jl}+\delta^{il}\delta^{jk}\right).\nonumber 
\end{align}

Doing so we obtain the effective action in Eq. (\ref{eq:FinalEffectiveAction}).
\begin{widetext}
\begin{align}
\mathcal{A}_{{\rm eff}} & =-\frac{\lambda}{\beta V}\sum_{p,\kappa}\mathcal{D}_{p,p+\kappa}\left[\bm{\xi}\cdot{\bf S}\left(\kappa\right)\right]\label{eq:FinalEffectiveAction}\\
 & +2\left(\frac{\lambda}{2\beta V}\right)^{2}\sum_{p,q}\sum_{\kappa,\kappa^{\prime}}\mathcal{D}_{p,q+\kappa}\mathcal{D}_{q,p+\kappa^{\prime}}\left[\bm{\xi}\cdot{\bf S}\left(\kappa\right)\right]\left[\bm{\xi}\cdot{\bf S}\left(\kappa^{\prime}\right)\right]\nonumber \\
 & +\left(\frac{\lambda}{2\beta V}\right)^{2}\sum_{p,q}\sum_{\kappa,\kappa^{\prime}}\Bigg\{\left[\mathcal{D}F\right]_{p,q+\kappa}\left[\mathcal{D}F\right]_{q,p+\kappa^{\prime}}-\xi^{2}\mathcal{D}_{p,q+\kappa}\mathcal{D}_{q,p+\kappa^{\prime}}\Bigg\}\left[{\bf S}\left(\kappa\right)\cdot{\bf S}\left(\kappa^{\prime}\right)\right]\nonumber 
\end{align}
\end{widetext}

with the operators $F$ and $\mathcal{D}$ being given by Eqs. (\ref{eq:FXI_F})
and (\ref{eq:TensorD}), respectively.

Note that the $1^{\mathrm{st}}$-order contribution (first term) vanishes
in the absence of the applied magnetic field. This contribution leads
to an effective gyromagnetic factor of the localized spins \citep{jenmac_ox1991,MyersEtAl_prb05,balcerzak_pss06,LiuFurdyna_2006,verkac2024phys}
which is linear in the magnetic field. The $2^{\mathrm{nd}}$-order
contribution yields two types of effective exchange coupling that
depend on the (effective) magnetic field ($\Xi$). This will be discussed
at length in Section \ref{sec:ESH}.

\subsection{Effective spin Hamiltonian of the MS in an external magnetic field}

In what follows we focus on the derivation of the effective Hamiltonian
for the localized spins under an external magnetic field, ignoring
the electrical disturbance.

In the presence of the splitting field ($\Xi\neq0$), $\mathcal{D}$
is given by Eq. (\ref{eq:TensorD}) and the $1^{\mathrm{st}}$ order
contribution to the effective action becomes

\begin{align*}
\mathcal{A}_{\mathrm{eff}}^{\left(1\right)} & =-\frac{1}{2}\frac{\lambda}{V}\sum_{k}\Bigg\{\frac{1}{\beta}\sum_{n}\left[\frac{1}{i\omega_{n}-E_{\bm{k}}^{-}}-\frac{1}{i\omega_{n}-E_{\bm{k}}^{+}}\right]\Bigg\}\left({\bf S}_{{\bf 0}}\cdot\bm{e}_{\xi}\right)
\end{align*}
where $\bm{e}_{\xi}=\bm{\xi}/\xi$, the verse of the magnetic field and ${\bf S}_{{\bf 0}}=\sum_{i}\bm{S}_{i}$.

Next, we make use of the usual technique to sum over the Matsubara
frequencies \citep{henfle_oxford2004} 
\begin{equation}
\frac{1}{\beta}\sum_{n\in\mathbb{Z}}g\left(i\omega_{n}\right)=\sum_{z_{0}\in\mathrm{poles}\left(g\right)}\underset{z=z_{0}}{\mathrm{Res}}\left[g\left(z\right)f_{\eta}\left(z\right)\right]\label{eq:MatsubaraSum}
\end{equation}
where $g\left(z\right)$ is supposed to be a holomorphic function
except at the poles $z=z_{0}$ and $f_{\eta}\left(z\right)$ is the
distribution function 
\[
f_{\eta}\left(z\right)=\left\{ \begin{array}{ll}
\frac{1}{e^{\beta z}+1}, & \mathrm{Fermi-Dirac}\,\left(\eta=-1\right),\quad\omega_{n}=\frac{\pi}{\beta}\left(2n+1\right),\\
\\
\frac{1}{e^{\beta z}-1}, & \mathrm{Bose-Einstein}\,\left(\eta=1\right),\quad\omega_{n}=\frac{2\pi n}{\beta}.
\end{array}\right.
\]

In particular, we have the general formula 
\begin{equation}
\frac{1}{\beta}\sum_{n}\frac{1}{\left(i\omega_{n}-\varepsilon\right)^{l}}=-\frac{\eta}{\left(l-1\right)!}\left(\frac{\partial^{l-1}f_{\eta}}{\partial\epsilon^{l-1}}\right)\left(\varepsilon\right).\label{eq:CauchyFormula}
\end{equation}

Consequently, the first contribution to the effective action becomes
\begin{align}
\mathcal{A}_{\mathrm{eff}}^{\left(1\right)} & =-\frac{\lambda}{2V}\sum_{\bm{k}}\left[f_{\mathrm{FD}}\left(E_{\bm{k}}^{+}\right)-f_{\mathrm{FD}}\left(E_{\bm{k}}^{-}\right)\right]\left({\bf S}_{{\bf 0}}\cdot\bm{e}_{\xi}\right).\label{eq:1stOrderAction}
\end{align}
This can also be rewritten as 
\[
\mathcal{A}_{\mathrm{eff}}^{\left(1\right)}=-\frac{\lambda}{8}\left(\frac{1}{V}\sum_{\bm{k}}\frac{\sinh\left(\beta\xi\right)}{\cosh\left(\beta E_{\bm{k}}^{+}\right)\cosh\left(\beta E_{\bm{k}}^{-}\right)}\right)\left({\bf S}_{{\bf 0}}\cdot\bm{e}_{\xi}\right)
\]
with 
\begin{equation}
E_{\bm{k}}^{\mu}=E_{\bm{k}}+\mu\Xi,\mu=\pm1\label{eq:FieldShiftedEnergy}
\end{equation}
The sum over $\bm{k}$ yields the population difference induced by
the effective magnetic field, or equivalently the polarization of
the band. Obviously, this effect vanishes for $\Xi=0$.

The calculation of the $2^{\mathrm{nd}}$-order contribution is more
involved and comprises several different contributions. It makes use
of the same technique for computing the various sums over Matsubara
frequencies. We also use the fact that the Fermi-Dirac distribution
function does not change under a translation by a bosonic frequency,
\emph{i.e.} $f_{\mathrm{FD}}\left(\epsilon+i\varpi_{m}\right)=f_{\mathrm{FD}}\left(\epsilon\right)$.

The final result for the $2^{\mathrm{nd}}$-order contribution to
the effective action then reads
\begin{widetext}
\begin{align}
\mathcal{A}_{{\rm eff}}^{\left(2\right)} & =\frac{1}{2}\left(\frac{\lambda}{2V}\right)^{2}\sum_{\bm{p},\bm{k}}\frac{1}{\beta}\sum_{m}\left[{\bf S}\left(-k\right)\cdot{\bf S}\left(k\right)\right]\Bigg\{\frac{f_{\mathrm{FD}}\left(E_{\bm{p}}^{-}\right)-f_{\mathrm{FD}}\left(E_{\bm{p}+\bm{k}}^{+}\right)}{i\varpi_{m}+E_{\bm{p}}-E_{\bm{p}+\bm{k}}-2\Xi}+\left(\Xi\longleftrightarrow-\Xi\right)\Bigg\}\label{eq:2ndOrder-Aeff_xi-noPP}\\
 & +\frac{1}{2}\left(\frac{\lambda}{2V}\right)^{2}\sum_{\bm{k},\bm{p}}\frac{1}{\beta}\sum_{m}\left[{\bf S}\left(-k\right)\cdot\bm{e}_{\xi}\right]\left[{\bf S}\left(k\right)\cdot\bm{e}_{\xi}\right]\nonumber \\
 & \Bigg\{\frac{f_{\mathrm{FD}}\left(E_{\bm{p}}^{-}\right)-f_{\mathrm{FD}}\left(E_{\bm{p}+\bm{k}}^{-}\right)}{i\varpi_{m}+E_{\bm{p}}-E_{\bm{p}+\bm{k}}}-\frac{f_{\mathrm{FD}}\left(E_{\bm{p}}^{-}\right)-f_{\mathrm{FD}}\left(E_{\bm{p}+\bm{k}}^{+}\right)}{i\varpi_{m}+E_{\bm{p}}-E_{\bm{p}+\bm{k}}-2\Xi}+\left(\Xi\longleftrightarrow-\Xi\right)\Bigg\}.\nonumber 
\end{align}
\end{widetext}

Note that the latter expression is an even function of the magnetic
field variable $\Xi$.

Therefore, the total effective action of the subsystem of localized
spins, in the presence of an external magnetic field, is given in
Eq. (\ref{eq:Aeff-noPP}). 

\section{\label{sec:Effective-exchange-couplings}Effective exchange couplings
in Fourier and direct space}

In this Appendix, we derive explicit analytical expressions of the
RKKY exchange couplings. These are needed in direct space for
the numerical solution of the Landau-Lifshitz equation and the study
of the dynamics of the effective spin model {[}see Section \ref{subsec:Magnetic-nutation_LLE}{]},
and in Fourier space for the analysis of the excitation modes and
their energy dispersion (see Section \ref{sec:precnut-freq}).

\subsection{\label{subsec:Exchange-coupling-inFS}Exchange coupling in Fourier
space}

In order to obtain explicit expressions for the exchange couplings
$\mathcal{J}_{\parallel}\left(\bm{k},\omega;\xi\right)$ and $\mathcal{J}_{\perp}\left(\bm{k},\omega;\xi\right)$
appearing in Eqs. (\ref{eq:Jperp}, \ref{eq:Jparal}), as functions
of the distance ($\bm{r}_{ij}$) between any pair of localized spins,
we need to compute the following generic integral 
\begin{equation}
I_{\mu\nu}\left(k\right)=\int\mathcal{D}p\,\frac{f\left(E_{\bm{p}}^{\mu}\right)-f\left(E_{\bm{p}+\bm{k}}^{\nu}\right)}{\omega+E_{\bm{p}}^{\mu}-E_{\bm{p}+\bm{k}}^{\nu}}.\label{eq:Imunu-1}
\end{equation}
where we have replaced the sum with an integral using $\mathcal{D}p=d^{3}p/\left(2\pi\right)^{3}$
; $E_{\bm{p}}^{\mu}=E_{\bm{p}}+\mu\Xi,\mu=\pm1$. We also introduce
the (modified) Fermi momentum and energy 
\begin{align*}
k_{\mathrm{F}}^{\mu} & =\sqrt{k_{\mathrm{F}}^{2}+\mu\frac{2m\Xi}{\hbar^{2}}}=k_{\mathrm{F}}\sqrt{1+\mu\frac{\Xi}{\epsilon_{\mathrm{F}}}},\quad\epsilon_{\mathrm{F}}=\frac{\hbar^{2}k_{\mathrm{F}}^{2}}{2m},\\
\epsilon_{\mathrm{F}}^{\mu} & =\frac{\hbar^{2}\left(k_{\mathrm{F}}^{\mu}\right)^{2}}{2m}=\epsilon_{\mathrm{F}}+\mu\Xi,\mu=\pm1\,\mbox{(or \ensuremath{\uparrow\downarrow})},
\end{align*}
together with the density of states at the Fermi energy 
\[
\rho_{\mathrm{F}}^{\mu}=\frac{mk_{\mathrm{F}}^{\mu}}{\hbar^{2}\pi^{2}}=\rho_{\mathrm{F}}\sqrt{1+\mu\frac{\Xi}{\epsilon_{\mathrm{F}}}}.
\]
The density of states of a three dimensional parabolic spectrum is
$\rho\left(\epsilon\right)=\frac{\left(2m\right)^{3/2}}{2\pi^{2}\hbar^{3}}\sqrt{\epsilon}$
and $\rho_{\mathrm{F}}=\rho\left(\epsilon_{\mathrm{F}}\right)=\frac{mk_{\mathrm{F}}}{\hbar^{2}\pi^{2}}$.
Finally, as usual, we introduce the dimensionless parameters 
\begin{align*}
x & =\frac{k}{k_{\mathrm{F}}^{\mu}},\quad\frac{k}{2k_{\mathrm{F}}^{\mu}}\equiv\tilde{k}_{\mu},\quad\omega_{\mu\nu}\left(\Xi\right)=\omega+\left(\mu-\nu\right)\Xi.
\end{align*}

Then, following the procedure described in many textbooks {[}see \emph{e.g.}
\citep{coleman2015}{]}, one obtains 
\begin{align}
I_{\mu\nu}\left(\bm{k},\omega;\Xi\right) & =\frac{\rho_{\mathrm{F}}^{\mu}}{4}\Lambda_{\mu\nu}^{-}\mathcal{F}\left(\frac{k}{2k_{\mathrm{F}}^{\mu}}\Lambda_{\mu\nu}^{-}\right)+\frac{\rho_{\mathrm{F}}^{\nu}}{4}\Lambda_{\mu\nu}^{+}\mathcal{F}\left(\frac{k}{2k_{\mathrm{F}}^{\nu}}\Lambda_{\mu\nu}^{+}\right)\label{eq:Imunu-general}
\end{align}
where 
\begin{equation}
\Lambda_{\mu\nu}^{\mp}\left(k,\omega;\Xi\right)=1\mp\frac{2m\omega_{\mu\nu}\left(\Xi\right)}{\hbar^{2}k^{2}}=1\mp\frac{\omega_{\mu\nu}\left(\Xi\right)}{E_{\bm{k}}}\label{eq:Lambda_munu}
\end{equation}
and

\[
\mathcal{F}\left(x\right)=\frac{1}{2}+\frac{1-x^{2}}{4x}\log\left[\frac{x+1}{x-1}\right]
\]
is the Lindhard function.

In case $\Xi=0$, $\omega_{\mu\nu}\equiv\omega,k^{\uparrow\downarrow}=k$,
and 
\begin{align*}
I_{\mu\nu}\left(\bm{k},\omega;0\right) & =\frac{\rho_{\mathrm{F}}}{4}\left(1-\frac{2m\omega}{\hbar^{2}k^{2}}\right)\mathcal{F}\left[\frac{k}{2k_{\mathrm{F}}}\left(1-\frac{2m\omega}{\hbar^{2}k^{2}}\right)\right]+\left(\omega\longrightarrow-\omega\right)\\
 & =\frac{\rho_{\mathrm{F}}}{4}\left[1+\frac{1}{4\tilde{k}}\left[1-\left(\tilde{k}-\frac{\tilde{\omega}}{\tilde{k}}\right)^{2}\right]\log\left[\frac{\left(\tilde{k}-\frac{\tilde{\omega}}{\tilde{k}}\right)+1}{\left(\tilde{k}-\frac{\tilde{\omega}}{\tilde{k}}\right)-1}\right]+\left(\omega\longrightarrow-\omega\right)\right]
\end{align*}
with $\tilde{\omega}=\omega/\left(4\epsilon_{\mathrm{F}}\right)$.

This is he result obtained by many authors in the absence of the external
magnetic field, see \emph{e.g.} the textbook \citep{coleman2015},
where the calculations are done for the dynamic spin susceptibility
with a factor of $2$ (from electron spin).

In the following, we will focus on the RKKY exchange coupling as a
function of space and this is given by the static limit of the couplings
(\ref{eq:Jperp}) and (\ref{eq:Jparal}). In this limit, (\ref{eq:Imunu-general}),
becomes ($\Lambda_{\mu\nu}^{\mp}\left(k,0;\xi\right)=1\mp\frac{2m\left(\mu-\nu\right)\xi}{\hbar^{2}k^{2}}$)
\begin{align}
I_{\mu\nu}\left(\bm{k},0;\Xi\right) & =\frac{\rho_{\mathrm{F}}^{\mu}}{4}\Lambda_{\mu\nu}^{-}\left(k,0;\Xi\right)\mathcal{F}\left(\frac{k}{2k_{\mathrm{F}}^{\mu}}\Lambda_{\mu\nu}^{-}\left(k,0;\Xi\right)\right)+\frac{\rho_{\mathrm{F}}^{\nu}}{4}\Lambda_{\mu\nu}^{+}\left(k,0;\Xi\right)\mathcal{F}\left(\frac{k}{2k_{\mathrm{F}}^{\nu}}\Lambda_{\mu\nu}^{+}\left(k,0;\Xi\right)\right).\label{eq:Imunu-general_static}
\end{align}

Accordingly, from Eq. (\ref{eq:Jparal}) we have

\begin{align}
\mathcal{J}_{\parallel}\left(\bm{k},0;\Xi\right) & =\frac{\lambda^{2}}{8}I_{--}\left(\bm{k},\omega;\Xi\right)+\left(\Xi\longrightarrow-\Xi\right)=\frac{\lambda^{2}}{8}\left[\frac{\rho_{\mathrm{F}}^{\uparrow}}{2}\mathcal{F}\left(\frac{k}{2k_{\mathrm{F}}^{\uparrow}}\right)+\frac{\rho_{\mathrm{F}}^{\downarrow}}{2}\mathcal{F}\left(\frac{k}{2k_{\mathrm{F}}^{\downarrow}}\right)\right].\label{eq:Jparal_mumu}
\end{align}

Now, we deal with the coupling $\mathcal{J}_{\perp}\left(\bm{k},\omega;\Xi\right)$.
Starting from Eq. (\ref{eq:Jperp}) and using again Eq. (\ref{eq:Imunu-general_static}),
we obtain ($\Lambda_{-+}^{\mp}\left(k,0;\Xi\right)=1\pm2\Xi/E_{\bm{k}}$)
and 
\[
I_{-+}\left(\bm{k},0;\Xi\right)=\frac{\rho_{\mathrm{F}}^{\mu}}{4}\Lambda_{-+}^{-}\left(k,0;\Xi\right)\mathcal{F}\left(\frac{k}{2p_{\mathrm{F}}^{\mu}}\Lambda_{-+}^{-}\left(k,0;\Xi\right)\right)+\frac{\rho_{\mathrm{F}}^{\nu}}{4}\Lambda_{-+}^{+}\left(k,0;\Xi\right)\mathcal{F}\left(\frac{k}{2p_{\mathrm{F}}^{\nu}}\Lambda_{-+}^{+}\left(k,0;\Xi\right)\right)
\]

\begin{align}
\mathcal{J}_{\perp}\left(\bm{k},0;\Xi\right) & =\frac{\lambda^{2}}{8}I_{-+}\left(\bm{k},0;\Xi\right)+\left(\Xi\longrightarrow-\Xi\right)\label{eq:Jperp-mu_v2}
\end{align}
or more explicitly 
\begin{equation}
\mathcal{J}_{\perp}\left(\bm{k},0;\Xi\right)=\frac{\lambda^{2}}{4}\frac{\rho_{\mathrm{F}}^{\uparrow}}{4}\left[1+\frac{\Xi}{2\epsilon_{\mathrm{F}}^{\uparrow}}\left(\frac{k}{2k_{\mathrm{F}}^{\uparrow}}\right)^{-2}\right]\mathcal{F}\left(\frac{k}{2p_{\mathrm{F}}^{\uparrow}}\left[1+\frac{\Xi}{2\epsilon_{\mathrm{F}}^{\uparrow}}\left(\frac{k}{2k_{\mathrm{F}}^{\uparrow}}\right)^{-2}\right]\right)+\left(\Xi\longrightarrow-\Xi\right).\label{eq:Jperp-mu_v3}
\end{equation}

\subsection{\label{subsec:Exchange-coupling-inDS}Exchange coupling in direct
space}

In order to derive the spatial dependence of the exchange couplings,
\emph{i.e} $\mathcal{J}_{\parallel}\left(\bm{r}_{ij}\right)$ and
$\mathcal{J}_{\perp}\left(\bm{r}_{ij}\right)$ appearing in the effective
Hamiltonian (\ref{eq:Ham_Eff}), we need to compute the inverse Fourier
transform of the couplings in Eqs. (\ref{eq:Jperp}, \ref{eq:Jparal}).
Using the definition 
\begin{align*}
\mathcal{J}_{\alpha}\left(\bm{r},0,\Xi\right) & =\int\mathcal{D}p\,e^{-i\bm{k}\cdot\bm{r}}\mathcal{J}_{\alpha}\left(\bm{k},0,\Xi\right)=\int_{0}^{\infty}\frac{kdk}{2\pi^{2}r}\mathcal{J}_{\alpha}\left(k,0,\Xi\right)\sin\left(kr\right)
\end{align*}
and following the standard procedure described, for example in Ref.
\onlinecite{kakehashi_springer12}, we obtain for the longitudinal
coupling

\begin{align}
\mathcal{J}_{\parallel}\left(r,\Xi\right) & =\frac{\lambda^{2}}{2\pi}\left[\frac{\rho_{\mathrm{F}}^{\uparrow}}{2}\left(k_{\mathrm{F}}^{\uparrow}\right)^{3}\Phi\left(2k_{\mathrm{F}}^{\uparrow}r\right)+\frac{\rho_{\mathrm{F}}^{\downarrow}}{2}\left(k_{\mathrm{F}}^{\downarrow}\right)^{3}\Phi\left(2k_{\mathrm{F}}^{\downarrow}r\right)\right]\label{eq:Jparal-DirectSpace}
\end{align}
with 
\[
\Phi\left(x\right)=\frac{\sin x-x\cos x}{x^{4}}.
\]
More explicitly, we have 
\begin{equation}
\mathcal{J}_{\parallel}\left(r,\Xi\right)=\frac{\lambda^{2}mk_{\mathrm{F}}^{4}}{2\pi^{3}\hbar^{2}}\frac{1}{2}\left[\left(1+\frac{\Xi}{\epsilon_{\mathrm{F}}}\right)^{2}\Phi\left(2k_{\mathrm{F}}^{\uparrow}r\right)+\left(1-\frac{\Xi}{\epsilon_{\mathrm{F}}}\right)^{2}\Phi\left(2k_{\mathrm{F}}^{\downarrow}r\right)\right].\label{eq:Jparal-DirectSpace_v2}
\end{equation}

For $\Xi=0$, this reduces to the well known result

\begin{equation}
\mathcal{J}_{\parallel}\left(r\right)=\frac{\lambda^{2}mk_{\mathrm{F}}^{4}}{2\pi^{3}\hbar^{2}}\Phi\left(2k_{\mathrm{F}}r\right).\label{eq:Jparal-DirectSpace_ZeroField}
\end{equation}

The calculation of the inverse Fourier transform of the transverse
exchange coupling (\ref{eq:Jperp-mu_v2}) is more involved. In Ref.
\cite{werwil_msp06} an expression for this coupling was given which
we have checked and reproduced below.
We have numerically computed the inverse Fourier transforms of the
couplings $\mathcal{J}_{\parallel}\left(r,\Xi\right)$ and $\mathcal{J}_{\perp}\left(r,\Xi\right)$
and checked that we recover the expressions in Eqs. (\ref{eq:Jparal_mumu},
\ref{eq:Jperp-mu_v3}). We have also checked their numerical values
against those obtained by the numerical studies of Refs. \cite{JalboutEtAl_apl02,SharmaEtAl_jmrt23}
for the specific case of ZnO of variable doping with Mn or Co.

One may resort to an expansion in terms of the field $\Xi$. However,
it turns out that the first contribution is of second order but the
corresponding expression is rather cumbersome and will not be used
here.

Using our notations and correcting a misprint, we rewrite the expression
for the inverse Fourier transform of the transverse coupling given
in Ref. \citep{werwil_msp06} as follows, 
\begin{align}
\mathcal{J}_{\perp}\left(r,0;\xi\right) & =\frac{\lambda^{2}m}{8\pi^{3}\hbar^{2}r^{4}}\Bigg\{\label{eq:ExchangeCoupling_ww}\\
 & +\left[\left(k_{\mathrm{F}}^{-}\right)^{2}r^{2}-2\right]\left(k_{\mathrm{F}}^{+}r\right)\cos\left(k_{\mathrm{F}}^{+}r\right)-\left(k_{\mathrm{F}}^{+}k_{\mathrm{F}}^{-}r^{2}-2\right)\left(\sqrt{k_{\mathrm{F}}^{+}k_{\mathrm{F}}^{-}}r\right)\cos\left(\sqrt{k_{\mathrm{F}}^{+}k_{\mathrm{F}}^{-}}r\right)\nonumber \\
 & +\left[\left(k_{\mathrm{F}}^{-}\right)^{2}r^{2}+2\right]\sin\left(k_{\mathrm{F}}^{+}r\right)-\left(k_{\mathrm{F}}^{+}k_{\mathrm{F}}^{-}r^{2}+2\right)\sin\left(\sqrt{k_{\mathrm{F}}^{+}k_{\mathrm{F}}^{-}}r\right)\nonumber \\
 & +k_{\mathrm{F}}^{+}k_{\mathrm{F}}^{-}r^{2}\left[\mathrm{Si}\left(k_{\mathrm{F}}^{+}r\right)-\mathrm{Si}\left(\sqrt{k_{\mathrm{F}}^{+}k_{\mathrm{F}}^{-}}r\right)\right]\Bigg\}\nonumber \\
 & +\frac{\lambda^{2}m}{4\pi^{3}\hbar^{2}r^{4}}\left[\left(k_{\mathrm{F}}^{+}k_{\mathrm{F}}^{-}r^{2}\right)\mathrm{H}_{0}\left(\sqrt{2k_{\mathrm{F}}^{+}k_{\mathrm{F}}^{-}}r\right)-r\sqrt{2k_{\mathrm{F}}^{+}k_{\mathrm{F}}^{-}}\,\mathrm{H}_{1}\left(\sqrt{2k_{\mathrm{F}}^{+}k_{\mathrm{F}}^{-}}r\right)\right]\nonumber 
\end{align}
where 
\[
\mathrm{Si}\left(z\right)=\intop_{0}^{z}dt\,\frac{\sin t}{t},
\]
and $\mathrm{H}_{n}\left(z\right),n=0,1$ is the Struve function that
is given by the solution of differential equation\citep{abrste_NY1970}
\[
z^{2}\frac{d^{2}y}{dz^{2}}+z\frac{dy}{dz}+\left(z^{2}-n^{2}\right)y=\frac{2}{\pi}\frac{z^{n+1}}{\pi\left(2n-1\right)!!}.
\]

\end{document}